\documentclass[prc,aps,floatfix,groupedaddress,amsmath,amssymb]{revtex4}
\usepackage{graphicx}
\usepackage{epsfig}
\usepackage{dcolumn}
\usepackage{bm}

\def\beq{\begin{equation}}
\def\eeq{\end{equation}}

\newcommand{\be}{\begin{eqnarray}}
\newcommand{\ee}{\end{eqnarray}}

\def\v#1{{\bf#1}}

\def\mH{\mathcal{H}}
\def\mK{\mathcal{K}}

\begin{document}
\thispagestyle{empty}
\vspace*{0.5 cm}
\begin{center}
{\bf Elementary excitations in homogeneous neutron star matter}
\\
\vspace*{1cm} {\bf M. Baldo and C. Ducoin}\\
\vspace*{.3cm}
{\it Dipartimento di Fisica, Universit\`a di Catania}\\
and \\
{\it INFN, Sezione di Catania}\\
{\it Via S. Sofia 64, I-95123, Catania, Italy} \\
\vspace*{.6cm}
\vspace*{1 cm}
\end{center}
{\bf ABSTRACT} \\
We study the collective density modes that can affect neutron-star thermodynamics
in the baryonic density range between nuclear saturation ($\rho_0$) and $3\rho_0$.
In this region, the expected constituents of neutron-star matter
are mainly neutrons, protons and electrons ($npe$ matter),
under the constraint of $\beta$ equilibrium.
The elementary excitations of this $npe$ medium 
are studied in the random-phase approximation framework.
We emphasize the effect of Coulomb interaction,
in particular the electron screening of the proton plasmon mode.
For the treatment of the nuclear interaction, we compare two modern Skyrme forces
and a microscopic approach.
The importance of the nucleon effective mass is observed.


\section{Introduction}

The evolution of neutron stars is in general determined by the microscopic processes that can occur in their
interior and in their crust. In particular, the long-term cooling by neutrino emission is strongly affected by
the detailed structure of the matter and the excitations it can sustain. The mean free path and emissivity of
neutrinos are directly related to the correlations present in neutron star matter, which in the standard models
is composed of neutrons, protons, electrons and muons~\cite{shap}.
The effects of collective excitations on neutrino emissivity have been extensively studied in the literature,
often with controversial results~\cite{Horowitz1,Horowitz2,Yako,Reddy,Kundu,Leinson1,Armen,Leinson2,Vosk}.
The plasmon excitations seem to play a relevant
role in quenching the contribution of the pair-breaking mode to neutrino emissivity~\cite{Leinson1},
since it obscures the possible effect of the Goldstone mode
that should be present in a neutral superfluid due to gauge invariance.
However, since neutrons, protons and electrons are coupled,
it is not clear to what extent the presence of the plasmon mode can affect
the whole set of collective excitations or the excitation spectrum at a qualitative level,
even for nonsuperfluid matter.
It has to be stressed that the plasmon is mainly an electron excitation,
since the electron screening on proton-proton Coulomb interaction is quite effective,
but the proton density oscillations can become relevant at not too low momentum.
Furthermore, some of these excitations are damped
and it is essential to estimate which actually are overdamped and therefore must be excluded.

Collective modes in asymmetric nuclear matter have been studied previously,
e.g. in Refs.~\cite{Haensel-NPA301, Matera-PRC49, Greco-PRC67}.
In the astrophysical context,
a study of the collective excitations in normal neutron star matter
on the basis of the relativistic mean-field method
has been presented in Ref.~\cite{Providencia-PRC74}.
In this article, we present a study of the excitation spectrum in homogeneous neutron star matter
under realistic conditions of proton fraction at different densities,
paying specific attention to the role of electron screening and to the damping of the modes.
Different nuclear models are used
to see the dependence of the spectrum on the nuclear interaction:
two modern Skyrme effective forces (SLy230a and NRAPR) are compared
with effective forces deduced from microscopic Brueckner-Hartree-Fock calculations.
In both types of approaches, the nucleons are considered nonrelativistic.
We treat the nucleonic part of the response function 
with the quantal random-phase approximation (RPA),
while the semiclassical Vlasov approximation is used for the electronic part (ultrarelativistic).
Since in this article we want to clarify the relevance of the nuclear and Coulomb forces,
the influence of pairing and the problem of neutrino emissivity are left to future works.

In Section~\ref{sec:formalism}, we briefly remind the RPA formalism
used to obtain the dispersion relations and strength functions.
Section~\ref{sec:coulcoup} is dedicated to the study of electron-proton coupling
through the pure Coulomb interaction,
to emphasize the suppression of the proton plasmon mode by electron screening.
The realistic situation, involving Coulomb and nuclear coupling
among neutrons, protons and electrons, is analyzed in Section~\ref{sec:coulnuc}.
Conclusions are exposed in Section~\ref{sec:conclusion}.


\section{RPA formalism}
\label{sec:formalism}

We will study the elementary excitations in homogeneous nuclear matter
in the RPA,
which amounts to considering correlations
only at the level of the one-particle$-$one-hole interaction.

Let us first consider a single fluid.
The RPA equations for the polarization propagator  $\Pi(q,\omega)$
is a function of the momentum $q$ and energy $\omega$:
\beq
\label{eq:RPA-1fluid}
\Pi(q,\omega) \, =\, \Pi_0(q,\omega)\, +\, v_{\rm{res}}(q) \Pi_0(q,\omega) \Pi(q,\omega) \;,
\eeq
where $v_{\rm{res}}$ is the residual (particle-hole) interaction
and $\Pi_0(q,\omega)$ is the free polarization propagator:
\beq \Pi_0(q,\omega) \,=\, \int \frac{d^3k}{(2\pi)^3} \int \frac{d\omega'}{2\pi i} G_0(\vert
\v{k}+\v{q}\vert,\omega+\omega'))G_0(\v{k},\omega') \;.
\eeq
\noindent $G_0(k,\omega)$ is the free single-particle Green' s function:
\beq
G_0(k,\omega) = \frac{\theta(k-k_{F})}{E-E_k + i\eta}\, +\,\frac{\theta(k_{F}-k)}{E-E_k - i\eta}
\eeq
\noindent where $E_k$ is the single-particle energy, $k_{{F}}$ the Fermi momentum
and $\theta(x)$ the Heaviside step function, which equals $1$ for $x > 0$ and zero otherwise.
The explicit expression for the free polarization in the nonrelativistic limit
can be found in textbooks (see for instance Ref.~\cite{Fetter-Walecka})
and is usually referred to as the Lindhard function.
Taking the low-momentum limit of this function,
we obtain the Vlasov (semiclassical) approximation,
whose accuracy increases for more relativistic particles~\cite{McOrist}.
In the present work, the nonrelativistic approximation is adopted for the nucleons,
while the ultrarelativistic and semiclassical (Vlasov) approximation
is taken for the polarization propagator of the electrons.

The RPA equations describing the motion of a multifluid system take the matrix form:
\beq
\label{eq:RPA-matrix}
\Pi^{ik}(q,\omega) \,
=\, \Pi_0^{i}(q,\omega)\left(\delta_{ik}\, + \,\, \Sigma_j\,\, v_{\rm res}^{ij}(q) \Pi^{jk}(q,\omega) \, \right) \;,
\eeq
which is an $N\times N$ matrix if $N$ particle species are present.
The branches of the dispersion relation
correspond to the poles of the response function.
Disregarding the imaginary part of the free polarization,
they are obtained by searching the zeros of the determinant $\Delta$ defined by:
\beq
\label{eq:Delta}
\Delta=\det\left[{\rm Re}\,(1-\mathbf{\Pi}_0\mathbf{v}_{\rm{res}})\right]
\eeq
where $\mathbf{\Pi}_0$ and $\mathbf{v}_{\rm{res}}$
are the free-polarization and residual-interaction matrices.
It is a general feature of the RPA equations that each fluid generates a pair of branches
if the interaction between its constituents is repulsive enough,
and the interfluid coupling is not too strong.
Typically, the pair of branches associated with the particle species $i$
is positioned on each side of the line $\omega=qV_{{F}i}$,
where $V_{{F}i}$ is the Fermi velocity.
This line corresponds to the singularity of the semiclassical Lindhard function
(proportional to the free response function $\Pi_0$ in the Vlasov approach):
$$ L(s_i)=2-s_i\ln\left|\frac{1+s_i}{1-s_i}\right| $$
with $s_i=\omega/(qV_{Fi})$.
The lower branch of each pair is strongly damped.
To highlight this structure and help identify the nature of the different branches,
the straight lines $\omega=qV_{{F}i}$ are drawn on all the multifluid dispersion relations
we present in this work (dotted lines).
To determine the relevant modes of excitations,
it is essential to take into account the imaginary part of the response function:
the quantity $-{\rm Im}\,\Pi^{ii}$
gives the strength function associated with the particle species $i$,
which shows peaks when collective modes are present.
For each excitation mode, the damping rate is measured by the width of the corresponding peak;
the mode is actually a resonance if this width is not too large.


\section{Pure Coulomb interaction}
\label{sec:coulcoup}

Before considering the whole coupling scheme among neutrons,
protons and electrons in neutron star matter conditions,
we found it quite instructive to consider the proton-electron Coulomb coupling alone,
disregarding the nuclear interaction.
This idealized case, which of course does not correspond to actual physical conditions,
will permit clarification of two points, also relevant for the general physical situation.
On one hand, it will make more transparent the role of electron screening on the proton dynamics and,
on the other hand, it will give clear indications on the strengths of the electron and proton motion
at the different excitation energies.
In this way we will also describe in some detail the method we are going to use throughout the article
for analyzing the excitation spectrum of homogeneous neutron star matter.

\begin{figure}[h]
\begin{center}
\includegraphics[width=0.8\textwidth]{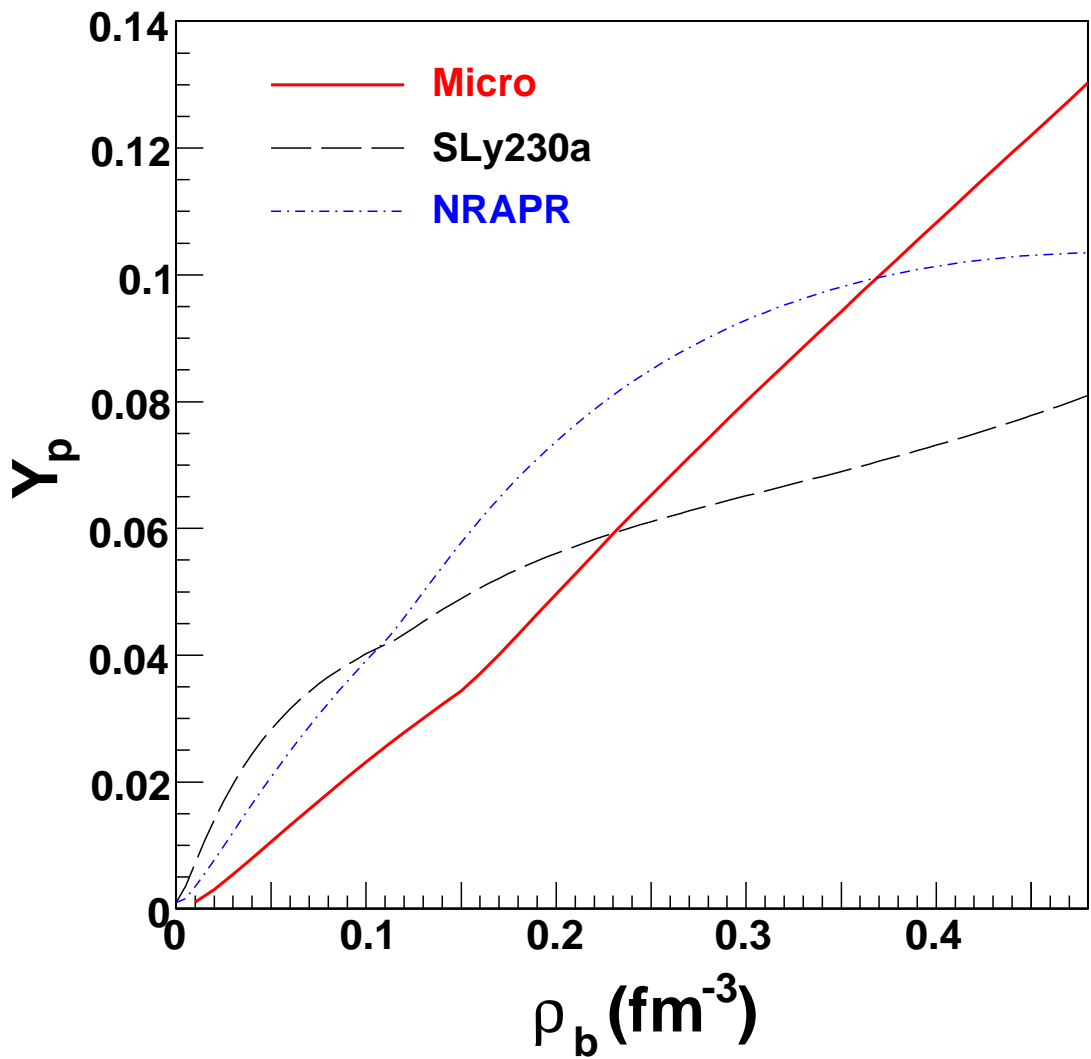}
\caption{
(Color online)
The proton fraction as a function of total baryon density in neutron star matter.
The $\beta$ equilibrium obtained from microscopic calculation is taken as a reference.
The curves obtained with the Skyrme-type forces are shown for comparison.
}
\label{fig1}
\end{center}
\end{figure}

As already mentioned in the Introduction, we are considering normal nuclear matter, leaving the case of
superfluid matter to a future work. The range of baryon density will be fixed between the saturation density
$\rho_0 = 0.16$ fm$^{-3}$ and $3\rho_0$, where indeed neutron star matter is expected to be homogeneous and not
affected by possible "exotic" components like hyperons. Of course the proton fraction has to be fixed to choose the proton density. This is determined by $\beta$ equilibrium and depends on the nuclear matter equation
of state (EOS). This will be taken from the microscopic calculations of Ref.~\cite{bbb,hans}, based on the
many-body theory within the Bethe-Brueckner-Goldstone (BBG) scheme. 
The calculated $\beta$ equilibrium takes into account the muon onset: 
however, for simplicity, muons and electrons will be treated in this work as a single ultrarelativistic fluid. 
We will work at zero temperature, but the results can be easily extended to finite temperature. 
The proton fraction as a function of baryon density is reported on
Fig.~\ref{fig1}. It is also given in Tab.~\ref{tab:fix} for selected values of the baryonic density,
together with the corresponding Fermi momenta and electron plasma frequency (at $\rho_e=\rho_p$). Throughout the article, we use the unit system $\hbar=c=1$.

\begin{table}[t]
\begin{center}
\begin{tabular}{ccccc}
\hline
$\rho$ & Y$_p$ & k$_{{F}n}$ & k$_{{F}p}$ & $\omega_{0e}$ \\[0.1cm]
\;[fm$^{-3}$]\; & \;[\%]\; & \;[MeV]\; & \;[MeV]\; & \;[MeV]\; \\[0.1cm]
\hline
0.16 & 3.7 & 327.2 & 110.5 & 6.15 \\
0.32 & 8.6 & 405.2 & 184.1 & 10.25 \\
0.48 & 13.0 & 456.2 & 242.3 & 13.48 \\
\hline
\end{tabular}
\end{center}
\caption{
Density dependence of the proton fraction, Fermi momenta and electron plasmon frequencies in neutron-star matter conditions.
The values given for the proton fraction $Y_{p}$ are obtained by microscopic calculation;
these values are adopted for the calculations performed with all models presented in this work.
}
\label{tab:fix}
\end{table}%

\begin{figure}[h]
\begin{center}
\includegraphics[width=0.8\textwidth]{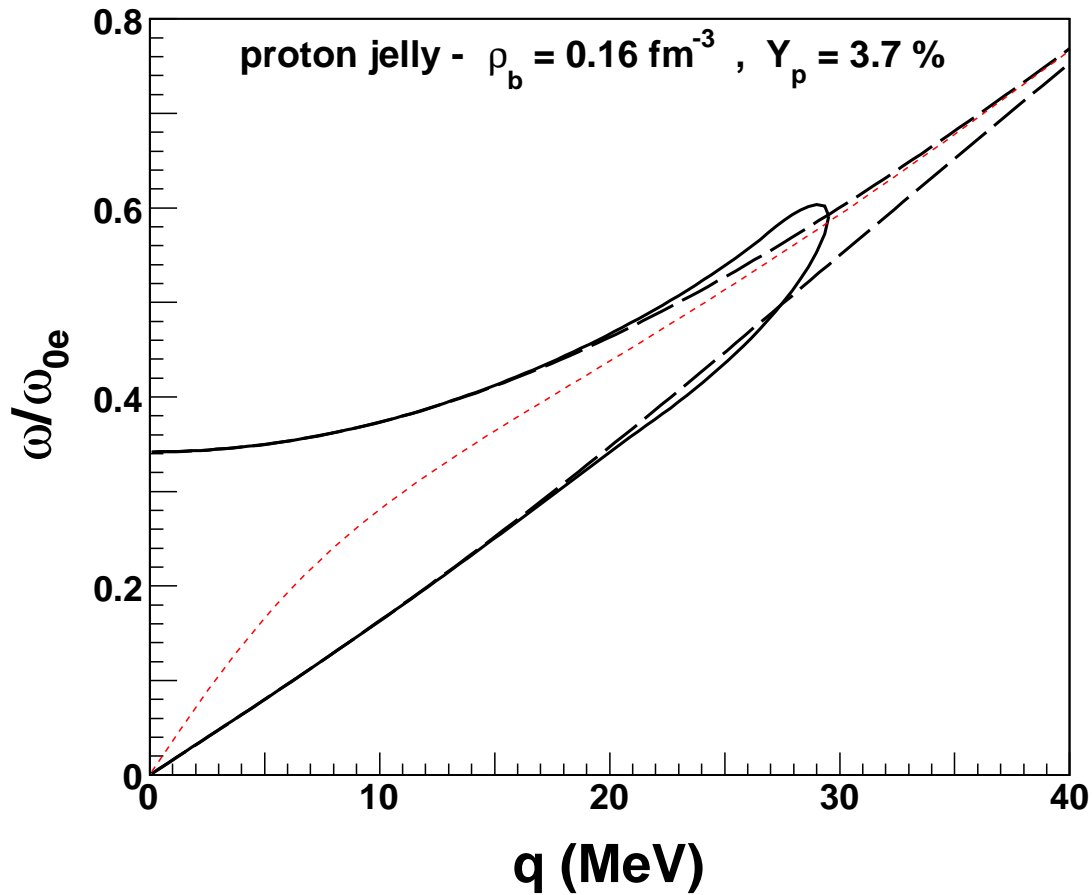}
\caption{
(Color online)
The proton dispersion relation in the jelly model.
The energy $\omega_{0e}$ is the electron plasmon energy at zero momentum.
The full line is obtained in the RPA scheme,
the long-dashed line corresponds to the Vlasov approximation.
The short-dashed line shows the effect of electron screening on the (Vlasov) proton plasmon branch.
}
\label{fig2}
\end{center}
\end{figure}

For a one-component plasma, i.e. a charged Fermi gas on a rigid uniform background (jelly model),
the RPA has been studied in detail,
both in the nonrelativistic~\cite{Fetter-Walecka} and relativistic~\cite{Jancovici} cases. 
In this case, the residual interaction is $v_{\rm{res}}=v_{\rm{c}}=4\pi e^2/q^2$. 
The excitation spectrum, i.e. the line in the energy-momentum plane 
where the real part of the polarization function~(\ref{eq:RPA-1fluid}) has poles, 
is known to have a "thumblike" shape. 
This is illustrated on Fig.~\ref{fig2} (full line), where the case of a proton gas at a density $\rho = \rho_0$ is considered. 
The upper branch appearing on this figure is called the plasmon excitation, 
which has a simple classical interpretation~\cite{Rax}. 
It is characterized by a finite energy at vanishing momentum: 
$\omega_0=V_{{F}}\sqrt{(4\pi e^2)N_0/3}$, 
where $V_{{F}}$ is the Fermi velocity and $N_0$ is the level density. 
The lower branch is actually strongly damped and does not correspond to a real excitation that is able to propagate. This point will be further elaborated in the sequel.
From the figure it is clear that in any case there is a maximum value of the momentum 
above which the spectrum stops and no excitation is possible. 
This is mainly a quantal effect, which is not present in the semiclassical approximation of RPA (Vlasov approximation). 
This is illustrated on the same Fig.~\ref{fig2}, where the RPA and the Vlasov approximation spectra are compared. 
Apart from the momentum cutoff, one can see that the spectra are quite similar.
The results for the relativistic case are analogous~\cite{Jancovici},
except that the "thumb" becomes thinner and thinner
as the particles at Fermi level become more and more relativistic~\cite{McOrist}.
In the ultrarelativistic case the two branches become extremely close near the cutoff,
and the difference  between RPA and Vlasov spectra appears smaller and smaller,
apart again for the presence of the momentum cutoff.

Having summarized the well-established results for the excitation spectrum with the pure Coulomb interaction
within the jelly model, let us consider the system of protons and electrons, treating  the two components on
equal footing ($pe$ model). In this case, the RPA equations are simply a $2\times 2$ system, where the coupling
is provided by the attractive Coulomb interaction between protons and electrons. The polarization propagator
$\Pi^{ik}$ is now a $2\times 2$ matrix: in Eq.~(\ref{eq:RPA-matrix}) the indexes $i, j, k$ stand for electrons
($e$) and protons ($p$). It has to be stressed that no exchange is included, i.e. only density-density
correlations are considered. Since electrons are expected to be faster than protons, they are able to follow
their motion and are more effective in screening the proton-proton interactions. The total spectrum is now
formed by four branches, but only one is expected to have the characteristic property of a finite excitation
energy at vanishing momentum: this will be identified with the plasmon mode of the proton-electron coupled
system. The original proton plasmon mode is now reduced to a soundlike mode, since, due to the electron
screening, the effective proton-proton interaction becomes of finite range (screened Coulomb interaction). This
effect is illustrated by the short-dashed line of Fig.~\ref{fig2}. This is also a well known phenomenon in
classical plasma~\cite{Rax}. The screening can be easily seen in the present case by solving
Eq.~(\ref{eq:RPA-matrix}) for the proton-proton polarization propagator. In the $pe$ model, we consider only the
Coulomb interaction, and have the relation $v_{\rm{c}}^{pp}=v_{\rm{c}}^{ee}=-v_{\rm{c}}^{pe}=v_{\rm{c}}$. After
a short algebra, one finds:
\beq
 \left( 1\, -\, \Pi_0^{p} \frac{v_{\rm c}}{1 - \Pi_0^e v_{\rm c} }\right) \Pi^{pp}\, =\, \Pi_0^{p}\;.
\eeq
\par
\begin{figure}[h]
\begin{center}
\includegraphics[width=0.8\textwidth]{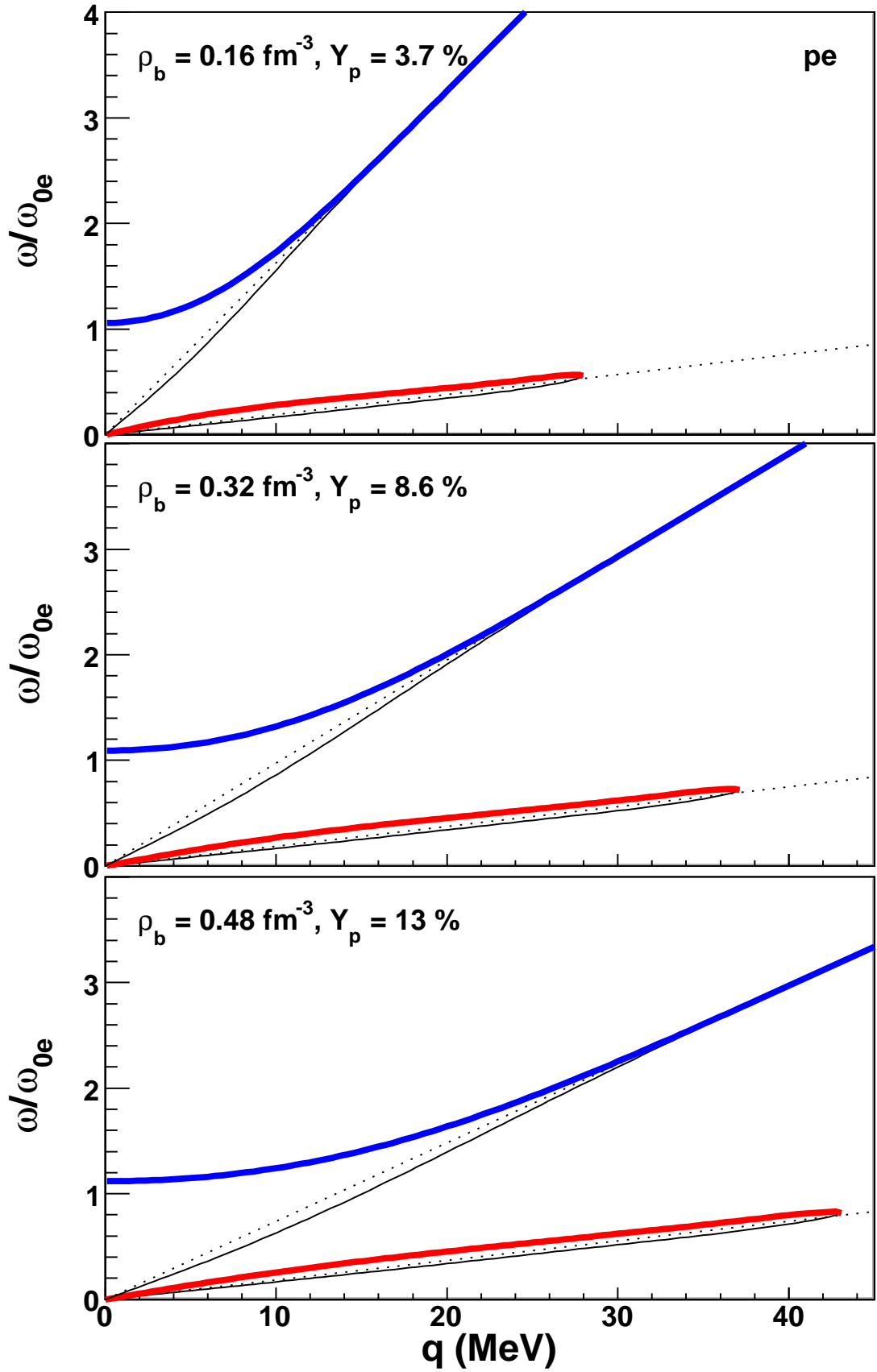}
\caption{
(Color online)
The four branches of the electron-proton system interacting by coulomb force only,
under neutron star matter conditions.
The energy $\omega_{0e}$ is the electron plasmon energy at zero momentum.
Only the thick branches are associated with collective modes,
while the thin ones are strongly damped.
The straight lines $\omega=qV_{{F}i}$ that structure the spectrum are indicated with dots.
}
\label{fig3}
\end{center}
\end{figure}

\noindent Since the unperturbed electron plasmon mode
is substantially higher than the corresponding proton one,
in the small $q$ limit one can expand the free electron polarization propagator
at zero frequency and one gets:
\beq
 \left( 1\, -\, \Pi_0^{p}\frac{4\pi e^2}{q^2 + q_{\rm c}^2}\right) \Pi^{pp}\, =\, \Pi_0^{p}\;.
 \label{eq:stat}
\eeq
This is equivalent to the RPA equation for a single fluid
with a screened (finite-range) Coulomb interaction.
For $q< q_{\rm{c}}$, the collective mode can be only a sound mode
with a typical linear dependence on momentum.
The inverse $\lambda = 1/q_{\rm{c}}$ is the electron screening length.
One finds that $ q_{\rm{c}}^2 = 3(\omega_{0e}/V_{{F}e})^2$,
where $\omega_{0e}$ is the electron gas plasmon frequency at zero momentum
and $V_{{F}e}$ the Fermi velocity.
In neutron star matter conditions, $q_{\rm{c}} \ll k_{{F}p}$
and the screening length is much larger than the average distance between protons.
Therefore, at increasing momenta the proton excitation
should merge in the proton plasmon mode with no screening.
The actual calculations confirm these expectations,
as illustrated on Fig.~\ref{fig3} for three proton densities corresponding to
total baryon densities $\rho_0$, 2$\rho_0$ and 3$\rho_0$.
The reported branches are obtained by searching, at given momentum,
the energies for which the determinant $\Delta$
of matrix ${\rm Re}\,[1-\mathbf{\Pi_0v_{\rm{res}}}]$ vanishes.
In the $pe$ model, $\Delta$ takes a simple expression:
\beq
 \Delta\, =\, 1\, -\, v_{\rm{c}} {\rm Re}\,[\Pi_0^{e}\, +\, \Pi_0^{p}]
 \label{eq:det}
\eeq
\noindent One observes that three branches go to zero at vanishing momentum.
The upper branch is the plasmon mode of the electron-proton system,
while the "thumblike" shape of the original proton branch in the uncoupled system
can be identified with the lower "loop".

\begin{figure}[h]
\begin{center}
\includegraphics[width=1.0\textwidth]{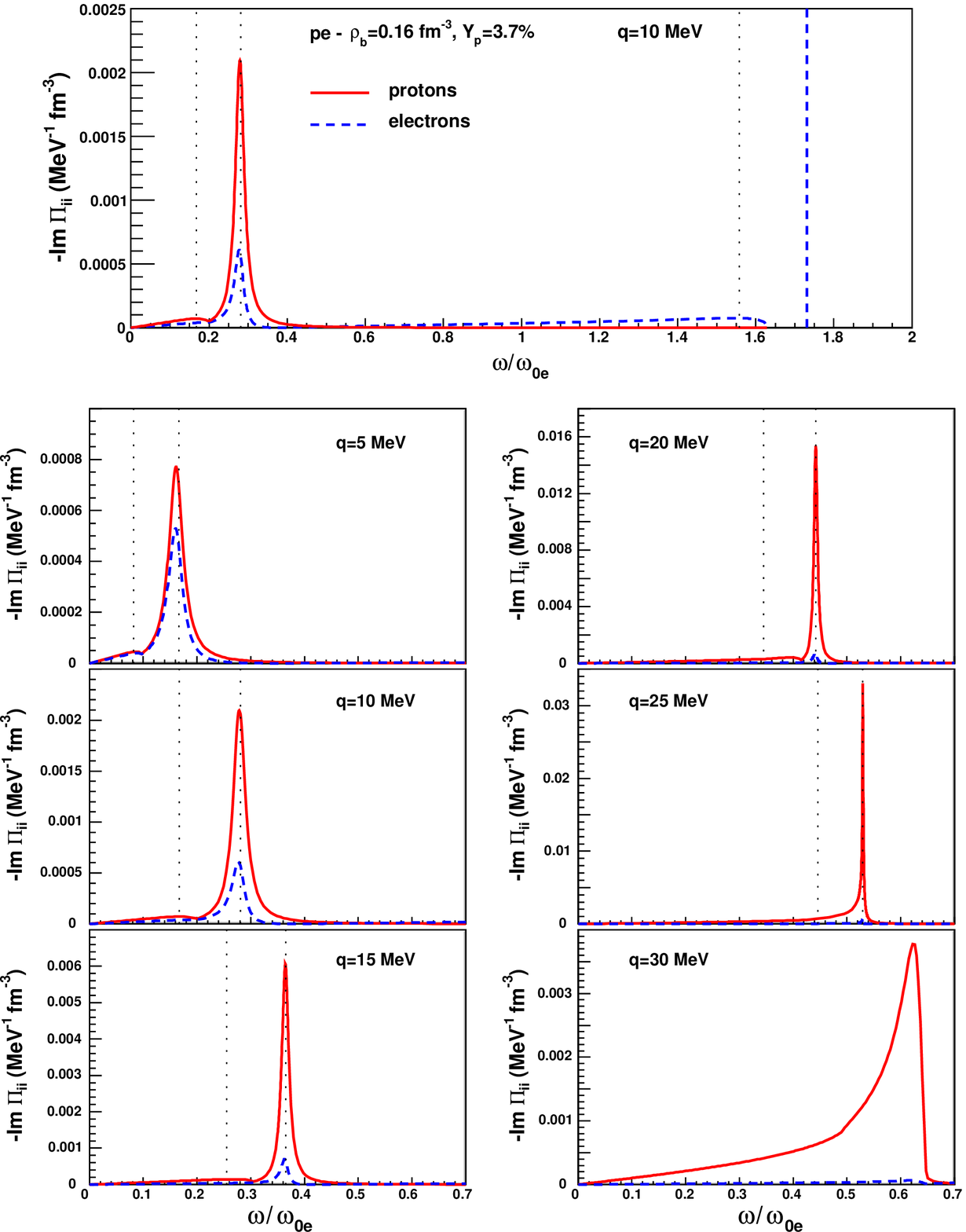}
\caption{
(Color online)
Electron and proton spectral functions at saturation density and for selected values of the momentum.
The vertical dotted lines show the energies
at which the determinant $\Delta$ of Eq.~(\ref{eq:det}) vanishes.
The plasmon mode coincides with the vertical line of highest energy.
The energy $\omega_{0e}$ is the electron plasmon energy at zero momentum.
Top: whole spectrum at $q$=10 MeV.
Bottom: evolution of the protonlike mode with the momentum.
The diagonal elements of the spectral function matrix are shown:
$-{\rm Im}\,\Pi^{pp}$ (full line), and
$-{\rm Im}\,\Pi^{ee}$ (dashed line).
}
\label{fig4}
\end{center}
\end{figure}

It has to be stressed that all branches, except the plasmon one, are actually damped. The polarization
propagators $\Pi_0$ that appear in Eq.~(\ref{eq:RPA-matrix}) are indeed in general complex. The best way to
estimate the damping rate of each branch is to calculate the strength function $-{\rm Im}\,\Pi^{ii}(q,\omega)$
(which is also a basic quantity for calculating the rate of many physical processes).
In the present case, we consider the proton and electron components ($i=p,e$).
At a given momentum $q$, as a function of $\omega$,
the strength function is expected to display peaks associated with the different branches,
their width giving a measure of the damping.
The electron and proton strength functions are reported on
Fig.~\ref{fig4} at saturation density for selected values of the momentum $q$.
One can see that peaks are present at the energies corresponding to the branches of Fig.~\ref{fig3},
even if some slight shifts occur.
Such shifts are expected, due to the contribution of the imaginary part.
The proton and electron strength functions display peaks approximately at the same energy.
The different heights at the peak maximum give a measure of the relative relevance of each components.
The plasmon, up to a certain momentum, has no width,
and the strength function presents a delta function singularity at the plasmon frequency.
To put in evidence the plasmon branch, we have reported just a sharp bar at the corresponding energy.
The relatively narrow peak at lower energy corresponds to the original proton plasmon in the uncoupled system.
As discussed above, in the coupled system, electron screening reduces this branch to a soundlike mode;
the coupling with electrons also produces a substantial damping.
For this mode, the proton component is larger, but electrons follow the proton oscillation,
which allows an important screening at low momentum.
However, as the momentum increases the screening is less effective, in agreement with the discussion above.
On Fig.~\ref{fig4}, we can observe that with increasing momentum (cases $q=5-25$ MeV)
the height of the peak increases and its width decreases.
This tendency ends with the branch associated with the mode:
at $q=30$ MeV, where the branch has stopped, the peak is still present but low and wide.
Let us finally comment the two other branches present in the spectrum (namely the lower branch of each pair).
In the strength functions, they are associated with bumps that can be hardly identified with collective modes:
they rather represent uncorrelated particle-hole excitations with a large spread in strength.
Indeed, once the peaks are subtracted, these bumps resemble the ones present in a free Fermi gas.
In any case, they can correspond only to strongly damped modes.
The behavior at higher density is quite similar.

\par
\begin{figure}[t]
\begin{center}
\includegraphics[width=0.8\textwidth]{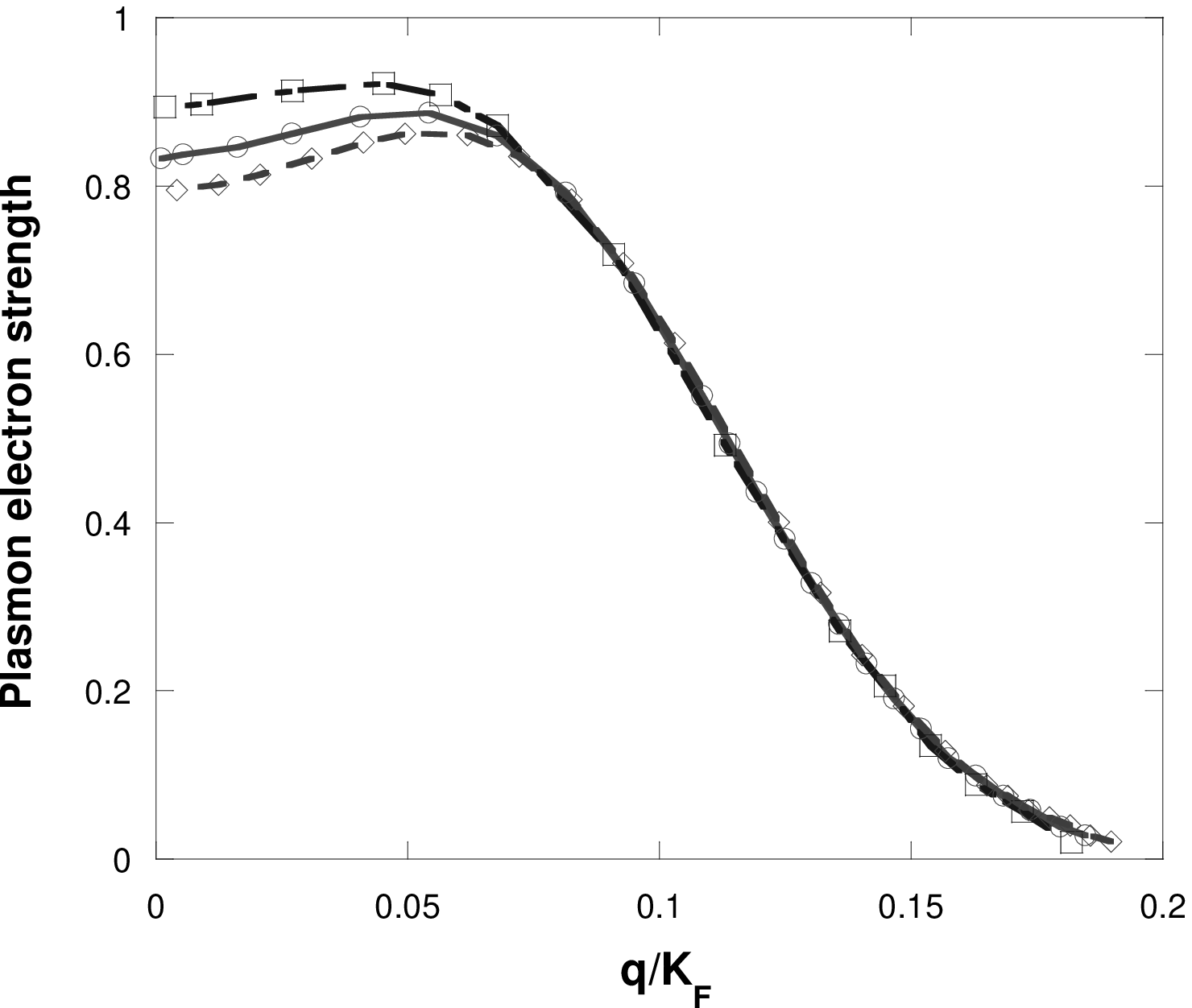}
\caption{
Fraction of the electron sum rule concentrated in the plasmon mode.
The three curves correspond to baryon density equal to
saturation density (diamonds), twice saturation density (circles)
and three times saturation density (squares).
}
\label{fig5}
\end{center}
\end{figure}

 An interesting check of the calculation is the fulfillment of the energy-weighted sum rules,
which in the present case can be written:
\beq
\label{eq:sumrule}
v_{\rm{c}}(q)\int d\omega \, {\rm Im}\,[\Pi^{ii}(q,\omega)]\,\omega \, =\, \frac{\pi}{2} (\omega_{0i})^2
\eeq
\noindent where $\omega_{0i}$ is the unperturbed plasmon frequency at zero momentum,
for electrons ($i = e$) and protons ($i =p$).
Notice that $\omega_{0p}/\omega_{0e} \approx k_{{F}}/M$,
where $k_{{F}}$ is the proton and electron Fermi momentum and $M$ the proton mass.
The sum rule is exact in the non-relativistic limit
and valid also in the relativistic case in the limit of low momenta and energy,
i.e. $q \ll k_{{F}}, \, \omega \ll E_{{F}}$, which corresponds to the Vlasov approximation.
The sum rules~(\ref{eq:sumrule}) are satisfied by the calculated strength functions
with an accuracy better than $10^{-3}$, which gives a check of the numerical accuracy.
The contribution of each peak to the corresponding sum rule
can be used to measure its relevance for many physical processes.
Let us consider in particular the plasmon mode of the coupled system.
Its contribution comes from a delta singularity,
and can be calculated by standard numerical methods.
The fraction of the sum rule for the electron component ($i = e$)
due to the plasmon excitation is reported on Fig.~\ref{fig5} as a function of momentum.
We see that the plasmon contribution decreases with increasing momentum.
One can also observe that the electron strength scales as $q/K_{{F}}$ for $q/K_{{F}}> 0.07$.
This can be easily understood by noticing that in this region
the plasmon energy $\omega \approx q$ (see Fig.~\ref{fig3}),
and then the electron polarization function depends only on $q/K_{{F}}$,
since no other scale is present.
For the proton component ($i = p$), one can find that the contribution of the plasmon to the corresponding sum rule
is never greater that few percentages and vanishingly small at the higher momenta.
This confirms that the plasmon mode of the $pe$ system is mainly an electron excitation.
On the contrary, the lower mode, corresponding to the sound excitation,
is mainly a (damped) proton excitation.

\begin{figure}[h]
\begin{center}
\includegraphics[width=0.8\textwidth]{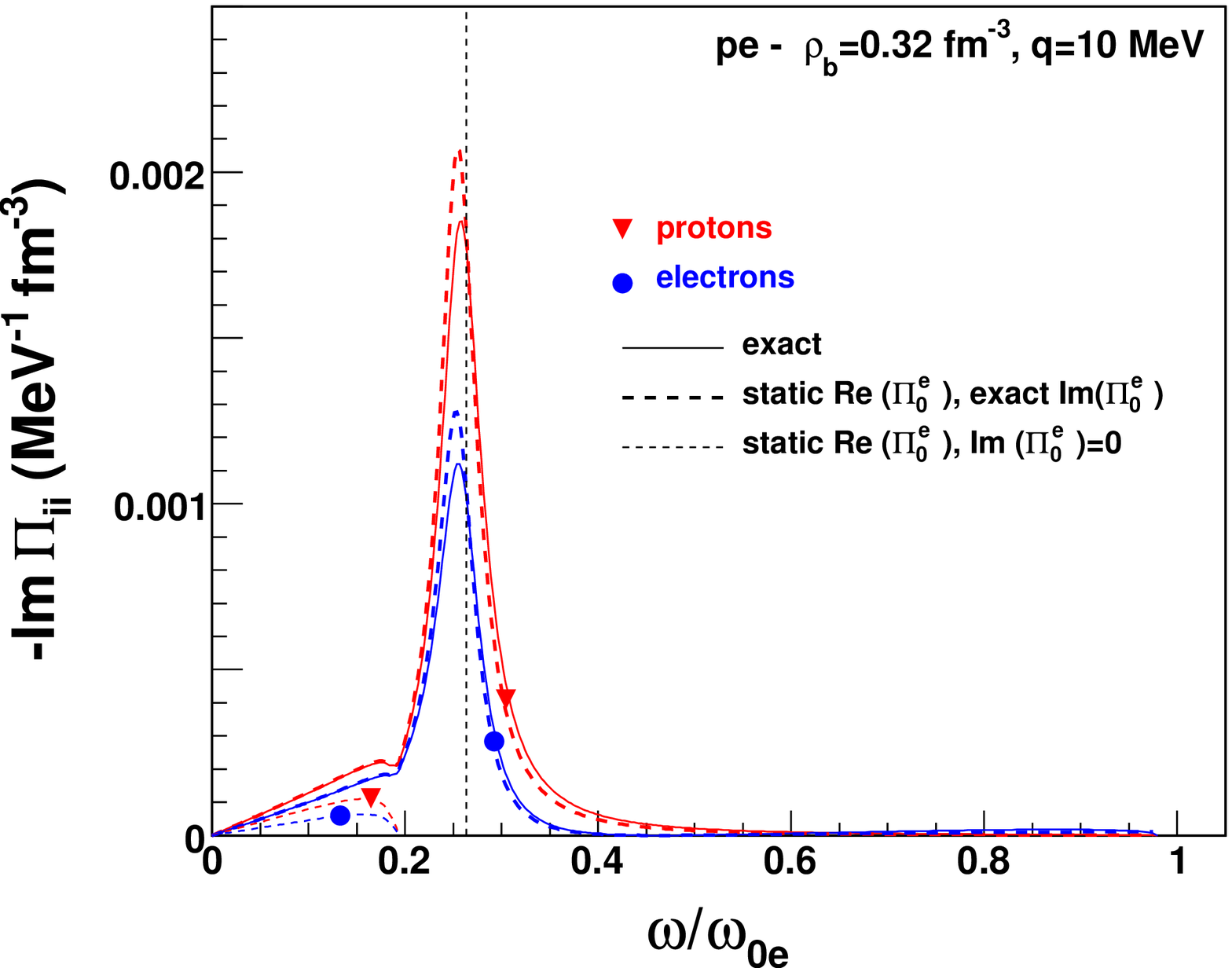}
\caption{
(Color online)
Effect of the static-electron approximation on the protonlike mode.
The energy $\omega_{0e}$ is the electron plasmon energy at zero momentum.
The diagonal elements of the spectral function matrix are shown:
$-{\rm Im}\,\Pi^{pp}$ (with triangular marker), and
$-{\rm Im}\,\Pi^{ee}$ (with round marker).
}
\label{fig6}
\end{center}
\end{figure}

The results we present in this article are obtained by solving explicitly Eq.~(\ref{eq:RPA-matrix}).
However, we found it interesting to check
the validity of the static-electron approximation of Eq.~(\ref{eq:stat}),
noticing that at high density the proton plasma frequency is a substantial fraction of the electron one.
In the strict static limit the proton strength function has a sharp peak (Dirac delta)
corresponding to the soundlike mode.
We found that the position of the peak is close to the maximum of the proton strength function corresponding to the damped mode obtained in the full calculation.
A better approximation can be obtained by treating only the real part of $\Pi_0^{e}$ in the static limit 
and leaving the imaginary part evaluated at the correct momentum.
In this case, the approximate proton strength function around the sound mode
is close to the exact one, as illustrated on Fig.~\ref{fig6}.
This is equally true for the electron strength function, which is also shown on the figure.
Of course, the static-electron approximation cannot be kept for the whole energy range
and in general one has to consider the exact strength function:
the full response function given by the matrix~(\ref{eq:RPA-matrix})
has to be calculated in its complete form.
This is what we will do in the rest of the article
and is in line with previous studies on neutrino interactions in neutron-star matter
~\cite{Horowitz1,Horowitz2,Yako,Reddy,Kundu,Leinson1,Armen,Leinson2,Vosk}.
Our discussion on the static approximation was intended
only to assess its validity and limitation.

\section{Elementary excitations in the realistic case.}
\label{sec:coulnuc}

\begin{table}[t]
\begin{center}
\begin{tabular}{ccccccccccc}
\hline
$\rho$ & Model &
v$_{\rm{res,b}}^{nn}$  & v$_{\rm{res,b}}^{pp}$  & v$_{\rm{res,b}}^{np}$ &
m$^*_n$/m & m$^*_p$/m &
V$_{{F}n}$ & V$_{{F}p}$ &
N$_{0n}$ & N$_{0p}$\\[0.1cm]
\;[fm$^{-3}$]\; &&
\;[MeV fm$^{3}$]\; & \;[MeV fm$^{3}$]\; & \; [MeV fm$^{3}$]\; &
&&
&&
\; [GeV$^{-1}$ fm$^{-3}$]\; & \;[GeV$^{-1}$ fm$^{-3}$]\; \\[0.1cm]
\hline
0.16 &
\begin{tabular}{l}
(Bare)\\
Micro\\
NRAPR\\
SLy230a\\
\end{tabular}
&
\begin{tabular}{c}
(0)\\
-46.2\\
47.1\\
-172.8\\
\end{tabular}
&
\begin{tabular}{c}
(0)\\
311.5\\
232.5\\
544.7\\
\end{tabular}
&
\begin{tabular}{c}
(0)\\
-394.1\\
-263.4\\
-117.9\\
\end{tabular}
&
\begin{tabular}{c}
(1)\\
0.926\\
0.806\\
0.544\\
\end{tabular}
&
\begin{tabular}{c}
(1)\\
0.856\\
0.610\\
0.969\\
\end{tabular}
&
\begin{tabular}{c}
(0.349)\\
0.376\\
0.432\\
0.640\\
\end{tabular}
&
\begin{tabular}{c}
(0.118)\\
0.137\\
0.193\\
0.121\\
\end{tabular}
&
\begin{tabular}{c}
(4.05)\\
3.75\\
3.27\\
2.21\\
\end{tabular}
&
\begin{tabular}{c}
(1.37)\\
1.17\\
0.83\\
1.33\\
\end{tabular}\\
\hline
0.32 &
\begin{tabular}{l}
(Bare)\\
Micro\\
NRAPR\\
SLy230a\\
\end{tabular}
&
\begin{tabular}{c}
(0)\\
168.4\\
252.1\\
88.3\\
\end{tabular}
&
\begin{tabular}{c}
(0)\\
284.3\\
381.1\\
606.1\\
\end{tabular}
&
\begin{tabular}{c}
(0)\\
-131.7\\
155.4\\
317.8\\
\end{tabular}
&
\begin{tabular}{c}
(1)\\
0.897\\
0.657\\
0.386\\
\end{tabular}
&
\begin{tabular}{c}
(1)\\
0.760\\
0.447\\
0.870\\
\end{tabular}
&
\begin{tabular}{c}
(0.432)\\
0.481\\
0.656\\
1.118\\
\end{tabular}
&
\begin{tabular}{c}
(0.196)\\
0.258\\
0.438\\
0.225\\
\end{tabular}
&
\begin{tabular}{c}
(5.02)\\
4.50\\
3.30\\
1.94\\
\end{tabular}
&
\begin{tabular}{c}
(2.28)\\
1.73\\
1.02\\
1.98\\
\end{tabular}\\
\hline
0.48 &
\begin{tabular}{l}
(Bare)\\
Micro\\
NRAPR\\
SLy230a\\
\end{tabular}
&
\begin{tabular}{c}
(0)\\
309.9\\
398.9\\
320.2\\
\end{tabular}
&
\begin{tabular}{c}
(0)\\
270.5\\
482.0\\
669.9\\
\end{tabular}
&
\begin{tabular}{c}
(0)\\
-5.2\\
459.6\\
588.8\\
\end{tabular}
&
\begin{tabular}{c}
(1)\\
0.951\\
0.543\\
0.306\\
\end{tabular}
&
\begin{tabular}{c}
(1)\\
0.848\\
0.357\\
0.746\\
\end{tabular}
&
\begin{tabular}{c}
(0.486)\\
0.510\\
0.894\\
1.587\\
\end{tabular}
&
\begin{tabular}{c}
(0.258)\\
0.304\\
0.722\\
0.345\\
\end{tabular}
&
\begin{tabular}{c}
(5.65)\\
5.37\\
3.07\\
1.73\\
\end{tabular}
&
\begin{tabular}{c}
(3.00)\\
2.54\\
1.07\\
2.24\\
\end{tabular}\\
\hline
\end{tabular}
\end{center}
\caption{
Density and model dependence of the residual interaction elements (bulk=momentum-independent part), effective masses, Fermi velocities and level densities. The bare values correspond to the free Fermi gas.
The proton fraction $Y_{p}$ is fixed by the $\beta$ equilibrium;
the $Y_{p}$ value obtained by microscopic calculation is taken for all models.
}
\label{tab:forces}
\end{table}%

The realistic physical situation in neutron stars
includes the effective nuclear particle-hole (p-h) interaction among neutrons and protons.
The effect of p-h correlations on neutrino mean free path has been analyzed in a few
articles~\cite{Burrows-Sawyer,Reddy}. Here we want to study in detail the spectral functions and the related
collective modes, including the dynamical electron screening. We compare the results from effective (Skyrme) and
microscopic nuclear models.
We restrict the analysis to density fluctuation modes, which are expected to include the most collective ones.
The particle-hole (or residual) interaction is derived in the Landau monopolar approximation.
It amounts to considering only the $F_0$ Landau parameter,
neglecting the parameters $F_L$ for $L > 0$.
This is equivalent to averaging the effective interaction over the Fermi surface
and is a reasonable approximation,
provided the nonlocality of the effective interaction is not too large.
This procedure not only simplifies the calculations, but also allows a more direct comparison
between the phenomenological treatment of the Skyrme type and the microscopic one,
where reliable EOS are available up to few times saturation density.
Furthermore, it makes a clear connection between the residual interaction
and the thermodynamical properties related to the (free) energy-density curvature.
Indeed, in the Landau monopolar approximation, the residual interaction reads~\cite{dmc-NPA809}:
\beq
\label{eq:vres-curv}
v_{\rm{res}}^{ij}=\frac{\delta^2\mH}{\delta\rho_i\delta\rho_j}-\delta_{ij}N_{0i}^{-1}\;,
\eeq
where $\mH$ is the energy density
and $N_{0i}=m^*_{i}k_{{F}i}/\pi^2$ is the level density, involving the effective mass $m^*_i$;
the indexes $(i,j)$ stand for neutrons ($n$) and protons ($p$).
An equivalent way to express it involves the single-particle energy levels:
\beq
\epsilon_i(k)=\frac{k^2}{2M}+U_i(k,\rho_n,\rho_p)\;,
\eeq
where single-particle potential $U_i(k)$ is obtained by the functional derivative
of the interaction energy density with respect to the single-particle density $\hat \rho_i$.
In the Landau monopolar approximation, we can identify the residual interaction
with the derivative of the single-particle potential at constant momentum, taken at Fermi level:
\beq
\label{eq:vres-derU}
v_{\rm{res}}^{ij}=\left(\frac{\delta U_i}{\delta\rho_j}(k_{{F}i},\rho_n,\rho_p)\right)_{k,\rho_{j^{\prime}\neq j}={\rm cst}}\;.
\eeq
It should be stressed that the nuclear p-h interaction in neutron-star matter condition
is not well known.
Only qualitative results can be derived,
which are relevant to understanding the possible structure of the spectra
and the connected phenomena that occur in neutron stars.
In the framework of effective models such as Skyrme-Hartree-Fock,
the residual interaction can be analytically deduced from the second derivative
of the energy-density functional.
We have obtained it applying directly Eq.~(\ref{eq:vres-curv}) to the Skyrme functional
(see Ref.~\cite{dmc-NPA809}).
However, this functional is still poorly constrained
for the high isospin asymmetries and densities involved in compact stars.
The adopted Skyrme parametrizations are SLy230a \cite{SLya} and NRAPR \cite{NRAPR}.
Both are modern Skyrme forces, whose fitting procedures
take into account microscopic calculations of the neutron-matter EOS.
However, important differences will be observed between these two models.
On the other hand, from the microscopic point of view,
we do not have an analytical expression for energy density $\mH(\rho_n,\rho_p)$.
The equation of state can be fitted by smooth functions,
but the consequences of the fits on the curvature are not well controlled,
especially at high asymmetry.
Furthermore, it is also important to note that the $k$ dependence of the single-particle potential $U_i$,
which is connected to the presence of an effective mass,
implies that even the interaction part of the energy-density curvature $\delta^2\mH_{\rm int}/\delta\rho_i^2$
diverges for low values of $\rho_i$, as explained in Appendix~\ref{ap:parabol}.
As a consequence, the isospin dependence of $\mH_{\rm int}$
cannot be treated in the parabolic approximation for high values of the asymmetry.
For these reasons, we have chosen to extract the microscopic residual interaction
rather from Eq.~(\ref{eq:vres-derU}),
performing numerical derivatives of $U_i(k_{{F}i},\rho_n,\rho_p)$.
The Brueckner-Hartree-Fock (BHF) potential is calculated following the scheme of Ref.~\cite{bbb,hans}, using the
Argonne v$_{18}$ two-body interaction with Urbana IX three-body force. The BHF calculations also provide the
effective masses to be used in the RPA equations. These effective masses must be carefully taken into account in
the derivation of the residual interaction (see Appendix~\ref{ap:parabol}). In all calculations the proton
fraction, i.e. the nuclear matter asymmetry, is taken from the corresponding microscopic calculations in neutron
stars. This  means that the calculations with Skyrme forces are not fully self-consistent. However, we do that to have a fair comparison between the Skyrme and microscopic results, since we want to single out the
structure of the excitation spectra under the same physical conditions, which should be the most realistic ones.

\begin{figure}[h]
\begin{center}
\includegraphics[width=1.0\textwidth]{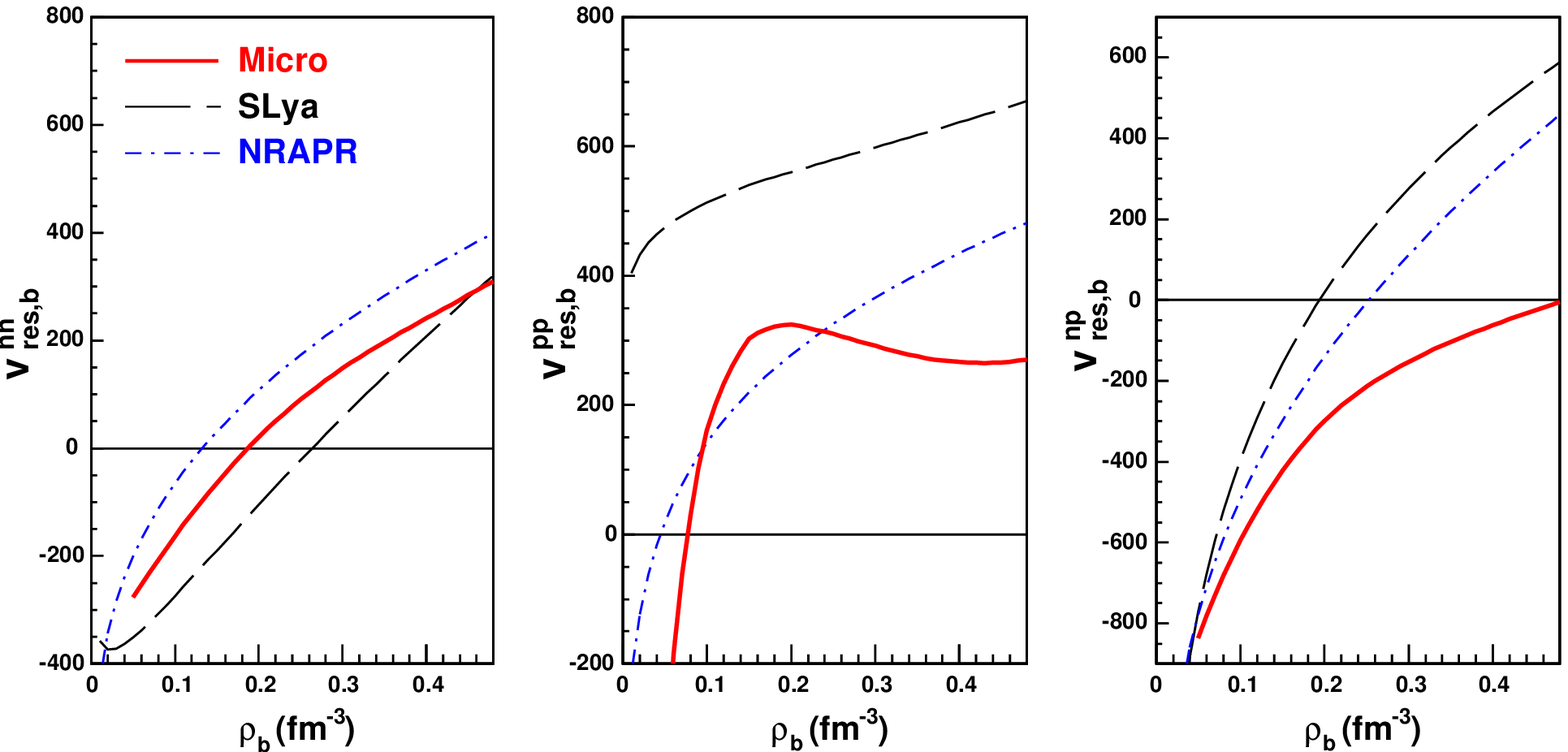}
\caption{
(Color online)
Elements of the residual-interaction matrix (bulk=momentum-independent part)
as a function of the baryonic density
in neutron star matter, for the different nuclear models: microscopic approach and Skyrme forces.
In the three cases, the proton fraction at $\beta$ equilibrium is fixed by the microscopic calculation.
}
\label{fig7}
\end{center}
\end{figure}

\begin{figure}[h]
\begin{center}
\includegraphics[width=1.0\textwidth]{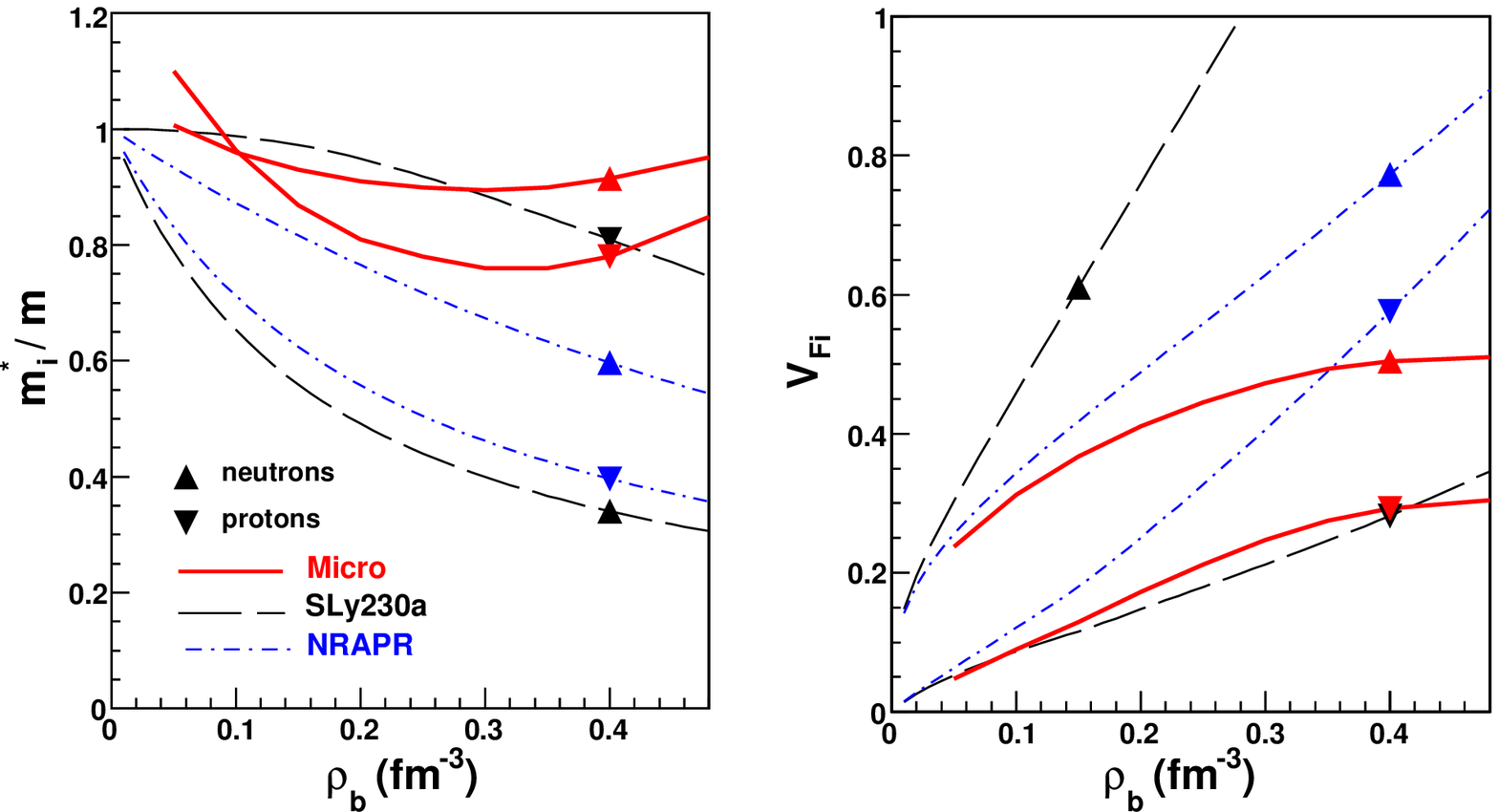}
\caption{
(Color online)
Effective mass (left) and Fermi velocity (right) as a function of the baryonic density
in neutron star matter, for the different nuclear models: microscopic approach and Skyrme forces.
In the three cases, the proton fraction at $\beta$ equilibrium is fixed by the microscopic calculation.
}
\label{fig8}
\end{center}
\end{figure}

The residual interaction elements $v_{\rm{res,b}}^{ij}$ obtained with the different models
are shown on Fig.~\ref{fig7} as a function of the baryonic density.
The subscript $b$ means "bulk":
this is the part of the residual interaction
that is independent of the transferred momentum $q$, contained in the nuclear part.
Note that, in the case of Skyrme forces,
the nuclear residual interaction contains a contribution proportional to $q^2$;
however, its effect is negligible at the relatively low $q$ values considered in this work.
In the following, the subscript $b$ will be omitted.
The effective masses and Fermi velocities are shown as a function of baryonic density
on Fig.~\ref{fig8}.
Furthermore, all the model-dependent quantities of relevance for the present study
are given in Tab.~\ref{tab:forces}, for the three selected values of the density we will consider.
We see that the Skyrme models SLy230a and NRAPR produce quite different effective p-h interactions.
Concerning the effective masses $m^*_i$, we can note the atypical behavior of SLy230a,
predicting a very low neutron effective mass
(in particular, in contrast with the NRAPR and microscopic case, it predicts $m^*_n<m^*_p$).
This results in a fast breakdown of the nonrelativistic description of the neutron fluid
at increasing density, as is obvious from the superluminal values of the Fermi velocity
that are reached even before $2\rho_0$.
Note that the NRAPR Fermi velocities also take large values as we approach $3\rho_0$.
Conversely, the quite high effective masses obtained in the microscopic framework
are more consistent with a nonrelativistic treatment of the nucleons.

We now turn to the study of collective density excitations including the nuclear force.
The particle-hole nuclear interaction results in the coupling of protons with neutrons, and
the possible appearance of neutronlike branches;
furthermore, the effective mass it produces modifies the density level and Fermi velocities,
affecting respectively the extension and orientation of the dispersion-relation branches
in the ($q,\omega$) plane.

\begin{figure}[h]
\begin{center}
\includegraphics[width=0.8\textwidth]{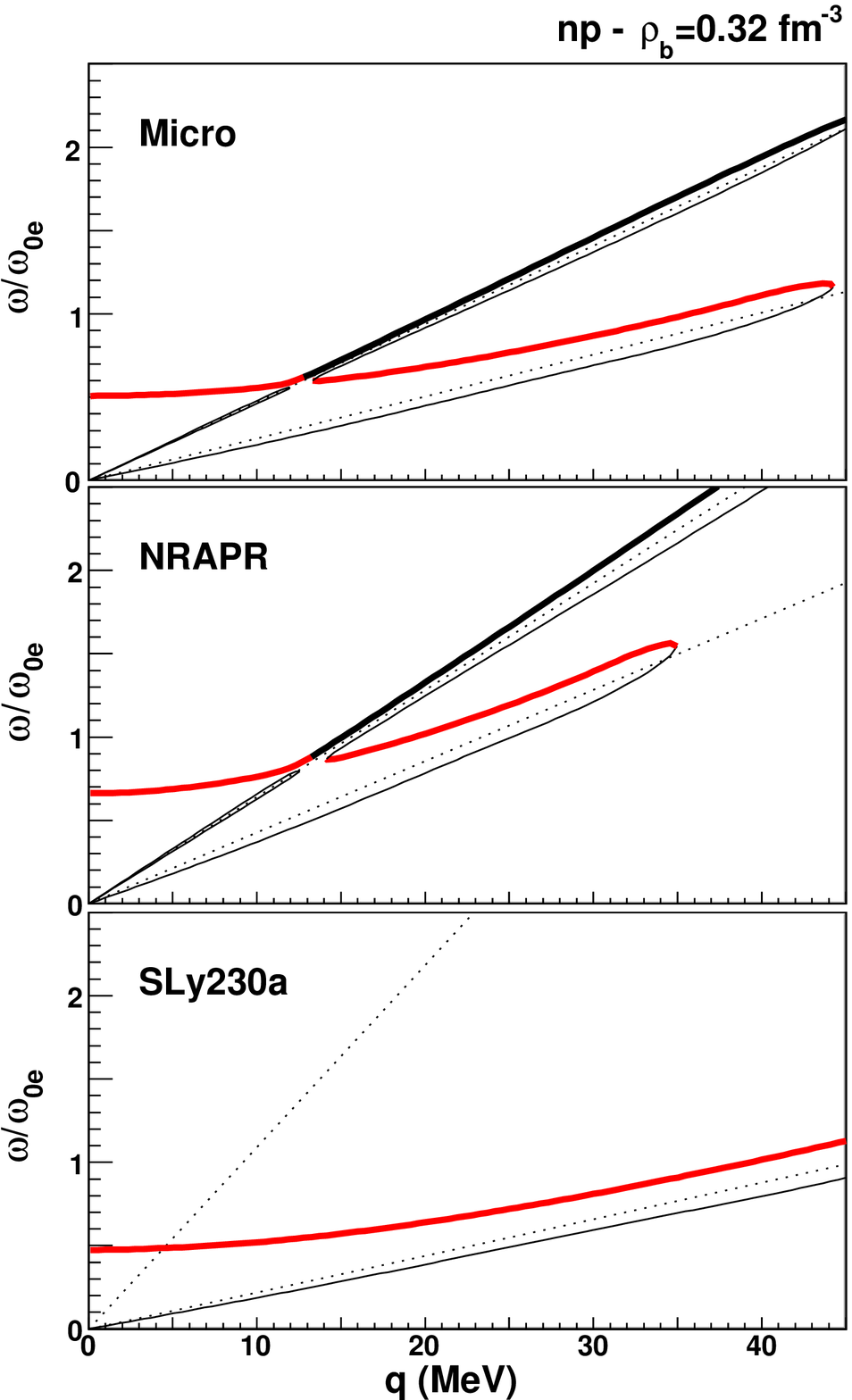}
\caption{
(Color online)
Dispersion relation in neutron$+$proton matter, at twice the saturation density
and with proton fraction under neutron star matter conditions, for the different nuclear models.
The energy $\omega_{0e}$ is the electron plasmon energy at zero momentum.
Only the thick branches are associated with collective modes,
while the thin ones are strongly damped.
The straight lines $\omega=qV_{{F}i}$ that structure the spectrum are indicated with dots.
}
\label{fig9}
\end{center}
\end{figure}

\begin{figure}[h]
\begin{center}
\includegraphics[width=1.0\textwidth]{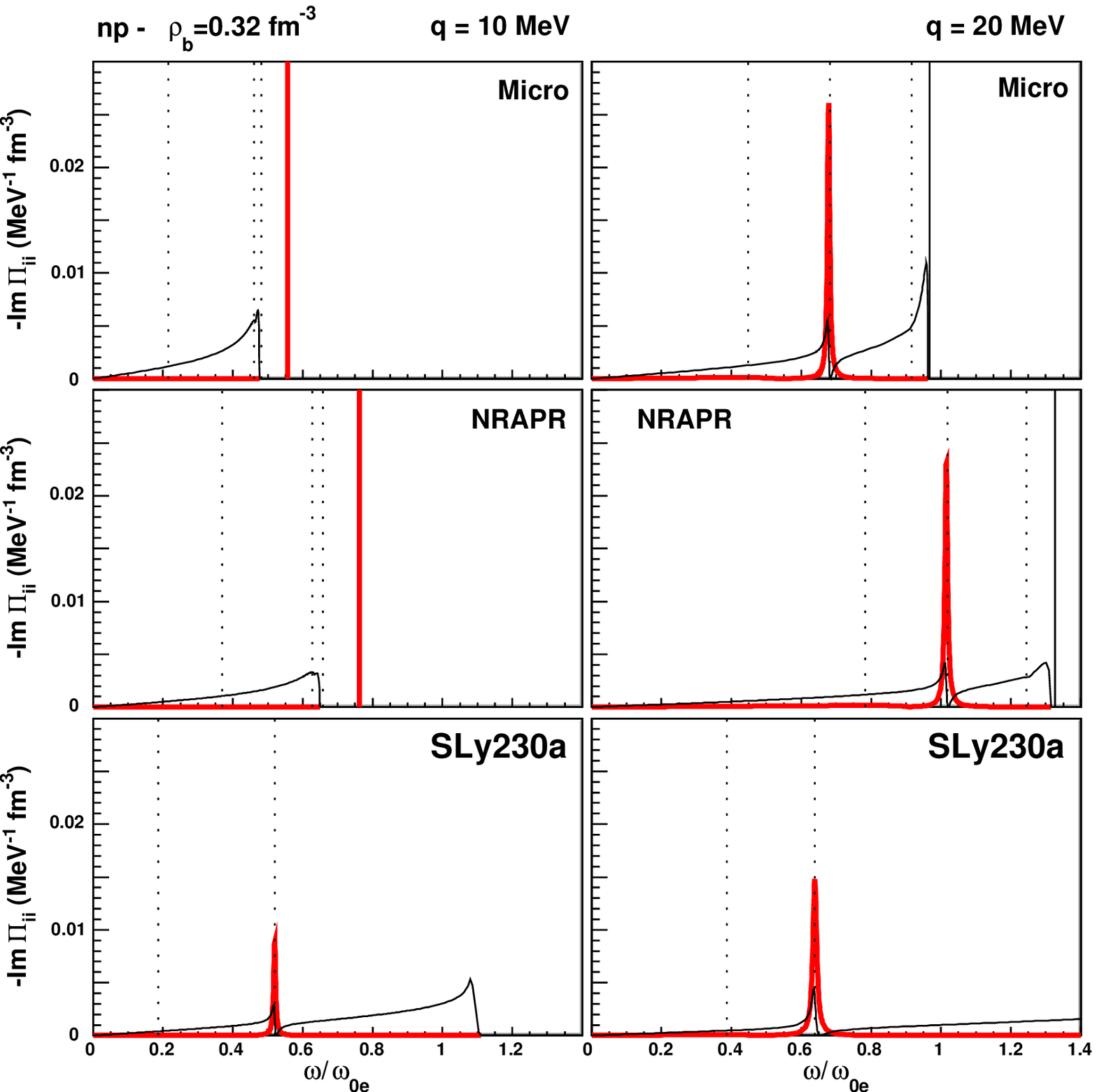}
\caption{
(Color online)
Spectral functions in neutron$+$proton matter for selected values of the momentum,
at twice the saturation density and with proton fraction under neutron star matter conditions, for the different nuclear models.
The energy $\omega_{0e}$ is the electron plasmon energy at zero momentum.
The diagonal elements of the spectral function matrix are shown:
$-{\rm Im}\,\Pi^{nn}$ (thin full line), and
$-{\rm Im}\,\Pi^{pp}$ (thick full line).
The vertical dotted lines show the energies
at which the determinant $\Delta$ of Eq.~(\ref{eq:Delta}) vanishes.
}
\label{fig10}
\end{center}
\end{figure}

Let us first study the simplified case
where only the neutron and proton dynamics are considered ($np$ model).
The electron dynamics is suppressed,
i.e. the electrons provide only a static uniform background of negative charge.
Of course this is not a realistic physical situation,
but it will help the interpretation of the results of the more complete and realistic calculations.
The protons interact also by the bare Coulomb potential and can display a plasmon mode.
However, the nuclear interaction, in particular the coupling between protons and neutrons,
can produce a shift and a width on the proton plasmon mode,
as well as other modes where the nuclear effective interaction is essential.
The dispersion-relation branches of this neutron+proton model
are displayed on Fig.~\ref{fig9} for the different nuclear models, at density $2\rho_0$.
In addition to the protonlike "thumb", two neutronlike branches appear
around the line $\omega=V_{{F}n}q$ in the NRAPR and microscopic case,
for which $v_{\rm res}^{nn}$ is strongly repulsive;
this is not the case with SLy230a, for which $v_{\rm res}^{nn}$ is weaker.
Since the line $\omega=V_{{F}n}q$ crosses the plasmon mode,
in presence of the neutronlike modes an exchange occurs around this crossing:
the upper branch, which at low $q$ corresponds to the plasmon mode,
turns into a neutronlike mode,
while the plasmon continues on the upper part of the thumb.
Note that in all three cases the protonlike mode is separated in two parts:
at low momentum, where the frequency of the plasmon branch is above $\omega=V_{{F}n}q$,
the plasmon mode is undamped (Dirac delta);
after the plasmon branch is crossed by the line $\omega=V_{{F}n}q$,
the imaginary part of the neutron response function induces a damping of the plasmon mode,
which becomes a peak of finite width.
This is shown by the strength functions displayed on Fig.~\ref{fig10}
for momenta $q=10$ and $20$ MeV.
We can see that, even when the plasmon mode is damped,
it corresponds to a neat resonance peak with a quite small width;
furthermore, it remains essentially a proton mode.
The behavior of the neutron strength function deserves some comments.
At $q=10$ MeV, it does not present any peak, even in the NRAPR and microscopic cases
where the relation dispersion presents neutronlike branches.
These branches do not correspond to physical modes in this part of the spectrum,
due to a strong damping.
However, at $q=20$ MeV where the highest branch is neutronlike,
the NRAPR and microscopic forces show a "genuine" undamped mode (Dirac delta)
near the edge of the neutron continuum.
This mode is due to a strongly repulsive neutron-neutron interaction.
A crucial role is also played by the effective masses.
For the SLy230a force the neutron effective mass is smaller than for the other two forces,
while for the proton effective masses it is the contrary.
For this force the neutron spectral function is expanded to a wider energy range,
while the proton one is shrunk to a smaller range.

\begin{figure}[h]
\begin{center}
\includegraphics[width=0.8\textwidth]{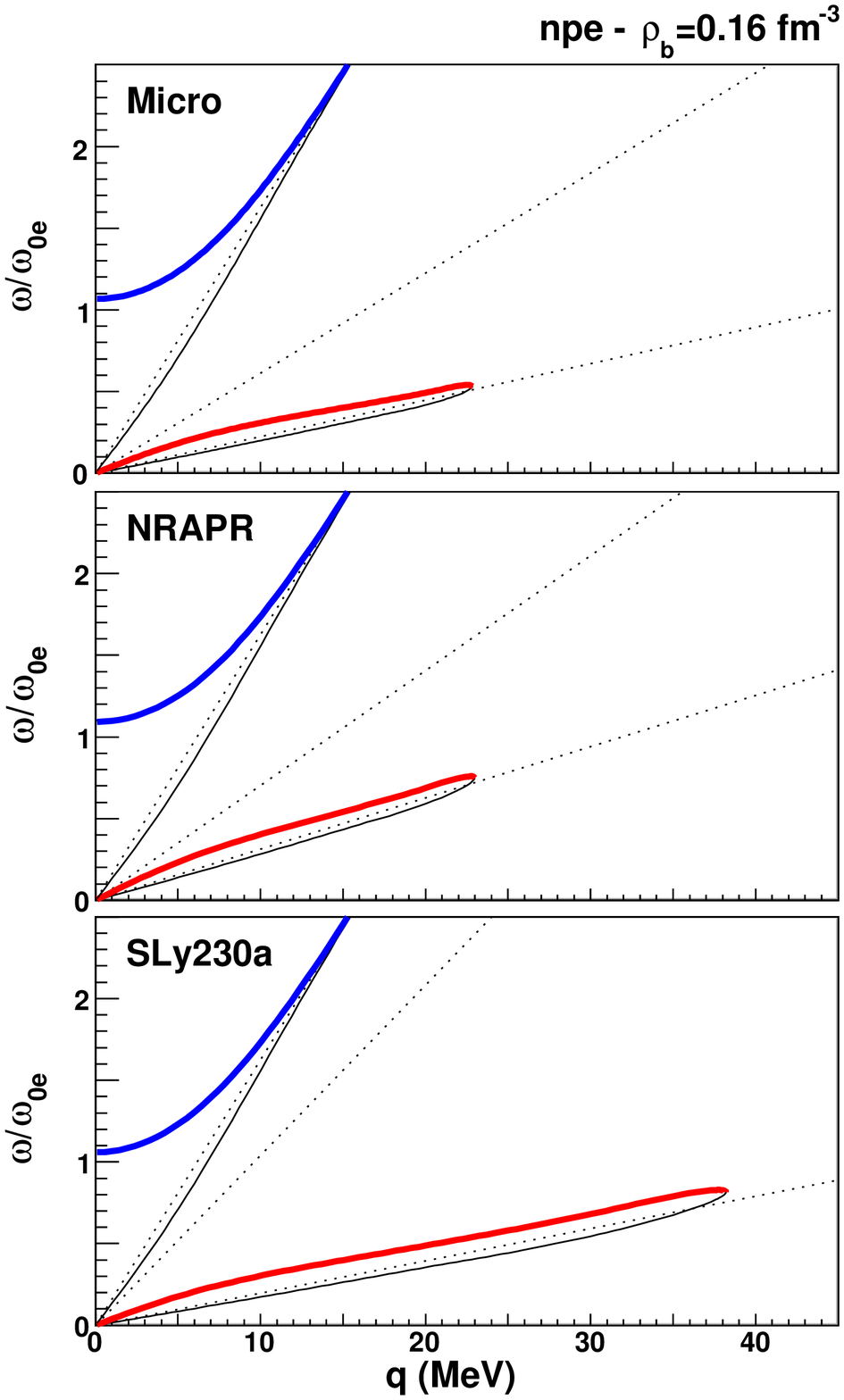}
\caption{
(Color online)
Dispersion relation in neutron$+$proton$+$electron matter, at saturation density
and with proton fraction under neutron star matter conditions, for the different nuclear models.
The energy $\omega_{0e}$ is the electron plasmon energy at zero momentum.
Only the thick branches are associated with collective modes,
while the thin ones are strongly damped.
The straight lines $\omega=qV_{{F}i}$ that structure the spectrum are indicated with dots.
}
\label{fig11}
\end{center}
\end{figure}

\begin{figure}[h]
\begin{center}
\includegraphics[width=0.8\textwidth]{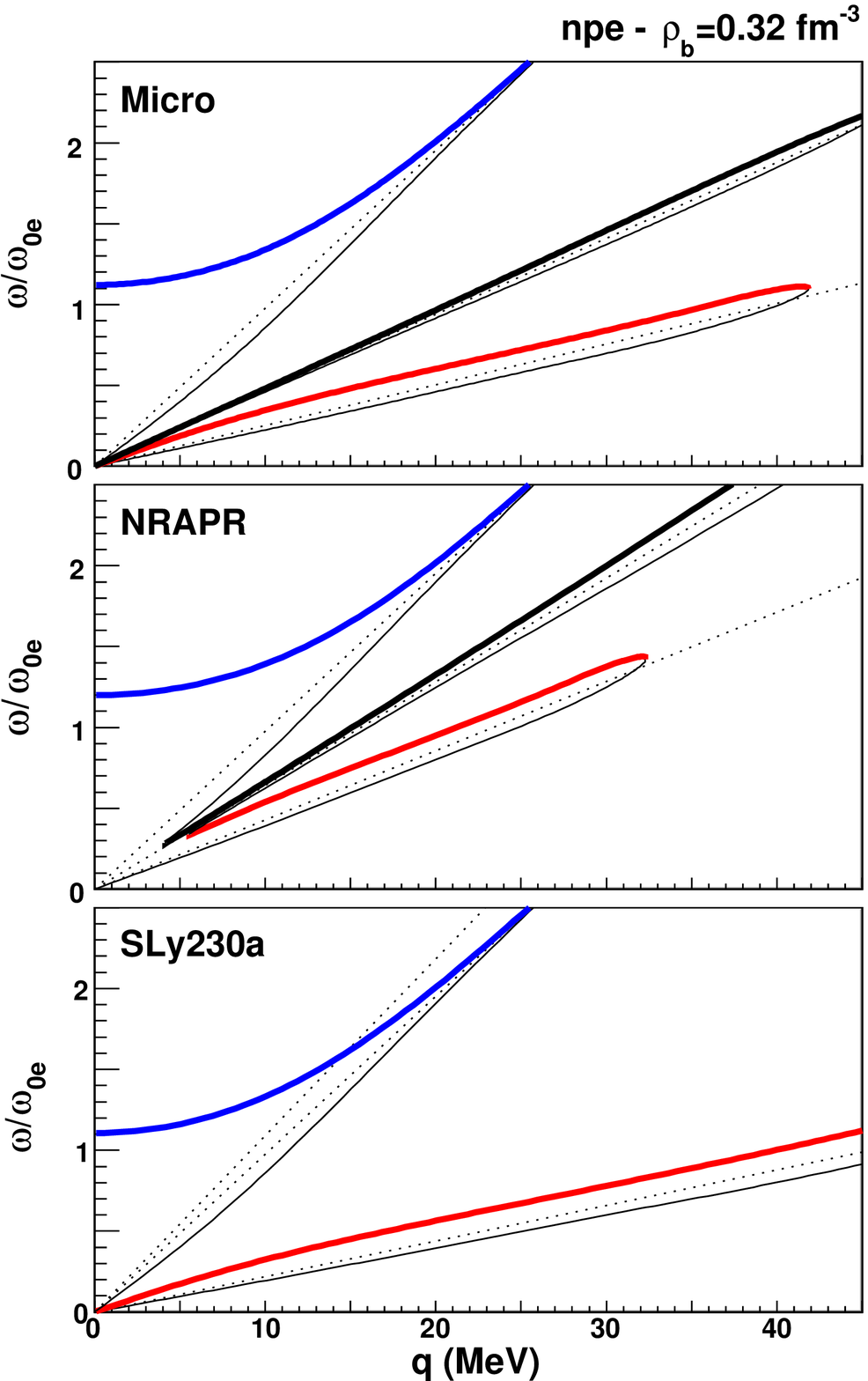}
\caption{
(Color online)
Dispersion relation in neutron$+$proton$+$electron matter, at twice the saturation density
and with proton fraction under neutron star matter conditions, for the different nuclear models.
The energy $\omega_{0e}$ is the electron plasmon energy at zero momentum.
Only the thick branches are associated with collective modes,
while the thin ones are strongly damped.
The straight lines $\omega=qV_{{F}i}$ that structure the spectrum are indicated with dots.
}
\label{fig12}
\end{center}
\end{figure}

\begin{figure}[h]
\begin{center}
\includegraphics[width=0.8\textwidth]{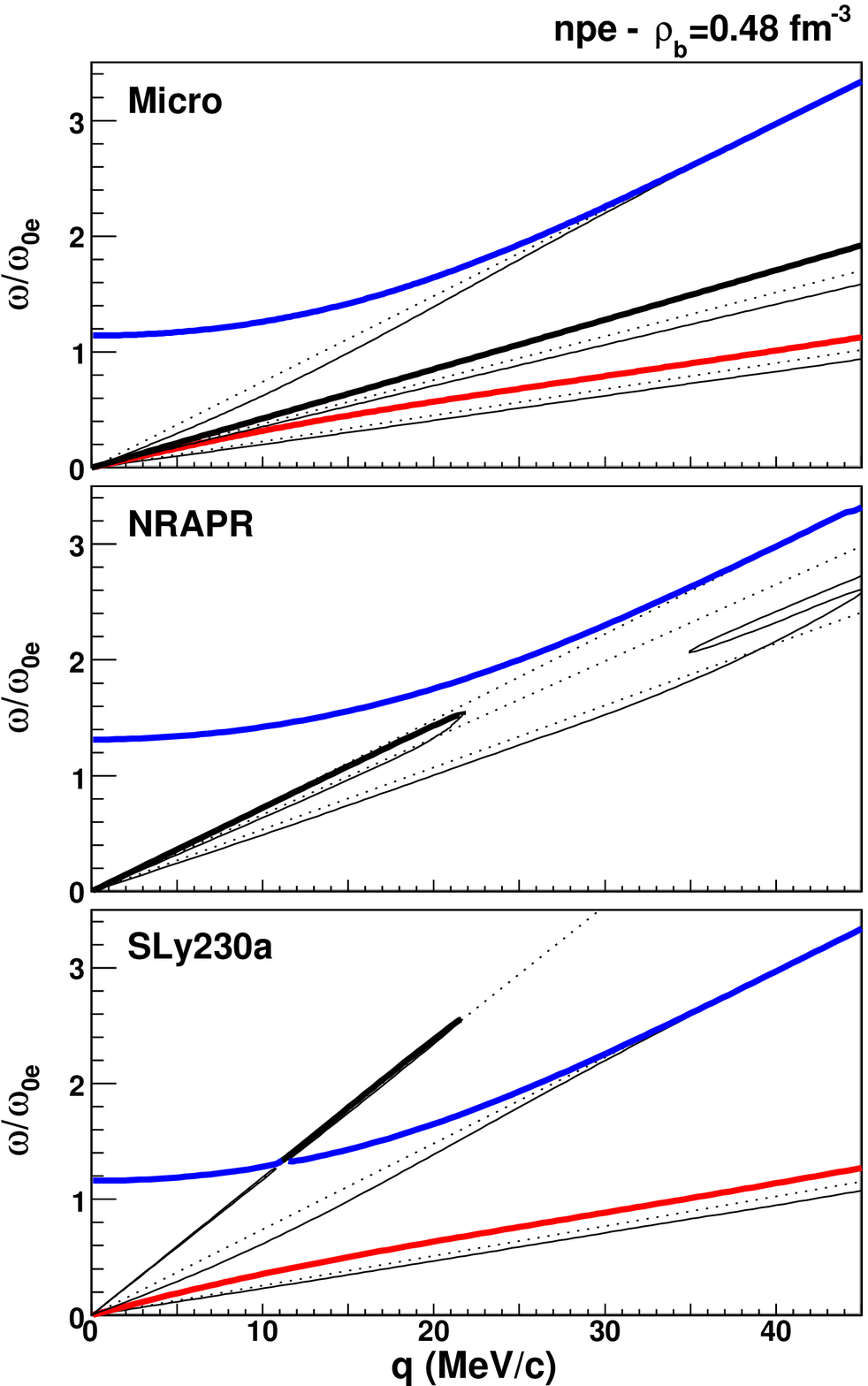}
\caption{
(Color online)
Dispersion relation in neutron$+$proton$+$electron matter, at three times the saturation density
and with proton fraction under neutron star matter conditions, for the different nuclear models.
The energy $\omega_{0e}$ is the electron plasmon energy at zero momentum.
Only the thick branches are associated with collective modes,
while the thin ones are strongly damped.
The straight lines $\omega=qV_{{F}i}$ that structure the spectrum are indicated with dots.
}
\label{fig13}
\end{center}
\end{figure}

All these findings are quite useful to understand the overall structure of the excitation spectra
in the realistic case when the electrons are dynamically included in the calculations.
The branches obtained for the full neutron+proton+electron ($npe$) system are shown
for the different models, for three values of the baryonic density:
$\rho_0$ (Fig.~\ref{fig11}),
$2\rho_0$ (Fig.~\ref{fig12}), and
$3\rho_0$ (Fig.~\ref{fig13}).
The proton plasmon is shifted down by screening to a soundlike mode,
as was the case without neutrons.
This shift in turn modifies the neutron-proton coupling, in general reducing it,
in particular for the SLy230a force, just because of the reduction of neutron phase space.
However, as we have seen, at higher momenta ($q>10$ MeV)
the screening disappears and the scheme is the same as without electrons,
except of course for the genuine plasmon mode at higher frequency
that is almost completely an electron excitation.
Proton-neutron excitations appear approximately decoupled from electron ones.
Concerning the nuclear interaction, comparing the different spectra
we observe that two kinds of quantities are crucial in determining their features
(as was already observed in the $np$ model):
the neutron-neutron residual interaction $v_{\rm{res}}^{nn}$,
and the effective masses.
Thus, the neutronlike branches appear when the neutron-neutron interaction
is sufficiently attractive: for all models, they are absent  at density $\rho_0$,
and appear at higher density.
On the other hand, the effective masses affect two important quantities:
the Fermi velocities $V_{{F}i}=k_{{F}i}/m^*_i$
and the level densities $N_{0i}=k_{{F}i}m^*_i/(\pi^2)$.
Since each pair of branches associated with a given particle species
tends to follow the corresponding line $\omega=qV_{{F}i}$,
neighboring Fermi velocities can lead to branch merging.
This is what happens with NRAPR at two and three times $\rho_0$
because of the high Fermi velocities of neutrons and protons,
in contrast with the microscopic case
(Figs.~\ref{fig12} and~\ref{fig13}).
The effect of the Fermi velocity is still more drastic with SLy230a,
where $V_{{F}n}$ takes unphysical superluminal values:
this leads to the crossing of the electron plasmon branch
by the line $\omega=qV_{{F}n}$.
In this case the plasmon is strongly damped through the Coulomb interaction with protons
and in turn through the coupling of the protons with neutrons by the nuclear effective interaction.
However, this situation is not realistic and only shows
the breakdown of the nonrelativistic treatment of the neutrons,
inconsistent with their low effective mass.
The second effect of $m^*_i$ concerns the level density:
indeed, the presence of a pair of branches associated with the particle species $i$
is favored by high values of $N_{0i}$.
This effect can be observed in different ways.
Let us consider, for instance, the NRAPR and microscopic spectra at density $2\rho_0$.
Despite a more repulsive interaction between protons,
the neutronlike branches extend to higher momenta due to the higher neutron level density.
Furthermore, although $v_{\rm{res}}^{pp}$ is more repulsive for NRAPR,
the protonlike branches have a greater extension in the microscopic case
because of the higher $m^*_p$ value (leading to a higher $N_{0p}$).
We can also compare the SLy230a and microscopic neutronlike branches at density $3\rho_0$;
although $v_{\rm{res}}^{nn}$ is more repulsive for SLy230a,
the corresponding branches are vanishing due to the low $m^*_n$ value,
in contrast with the microscopic case.

\begin{figure}[h]
\begin{center}
\includegraphics[width=1.0\textwidth]{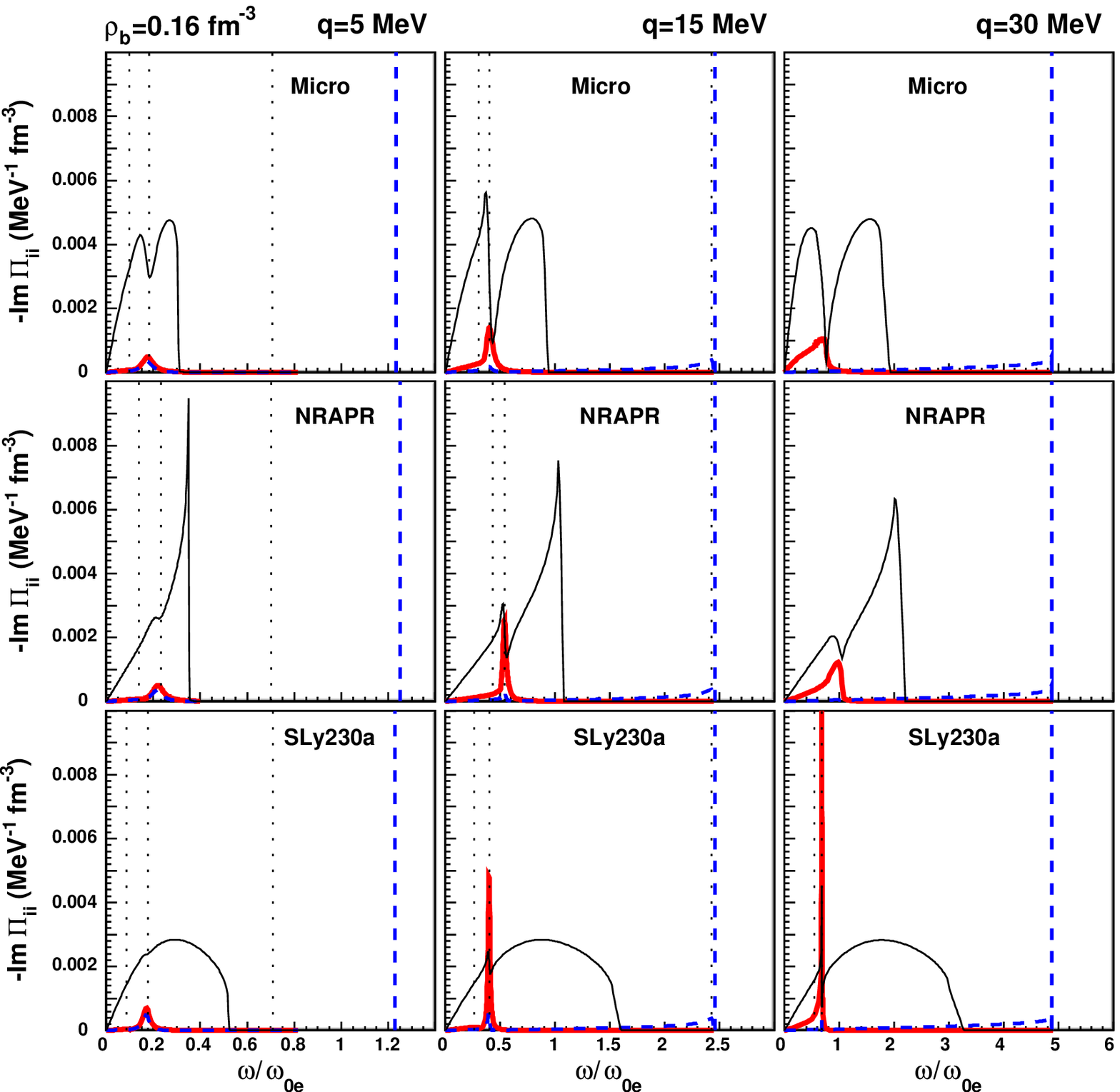}
\caption{
(Color online)
Spectral functions in neutron$+$proton$+$electron matter for selected values of the momentum,
at saturation density and with proton fraction under neutron star matter conditions, for the different nuclear models.
The energy $\omega_{0e}$ is the electron plasmon energy at zero momentum.
The diagonal elements of the spectral function matrix are shown:
$-{\rm Im}\,\Pi^{nn}$ (thin full line),
$-{\rm Im}\,\Pi^{pp}$ (thick full line), and
$-{\rm Im}\,\Pi^{ee}$ (dashed line).
The vertical dotted lines show the energies
at which the determinant $\Delta$ of Eq.~(\ref{eq:Delta}) vanishes.
}
\label{fig14}
\end{center}
\end{figure}

\begin{figure}[h]
\begin{center}
\includegraphics[width=1.0\textwidth]{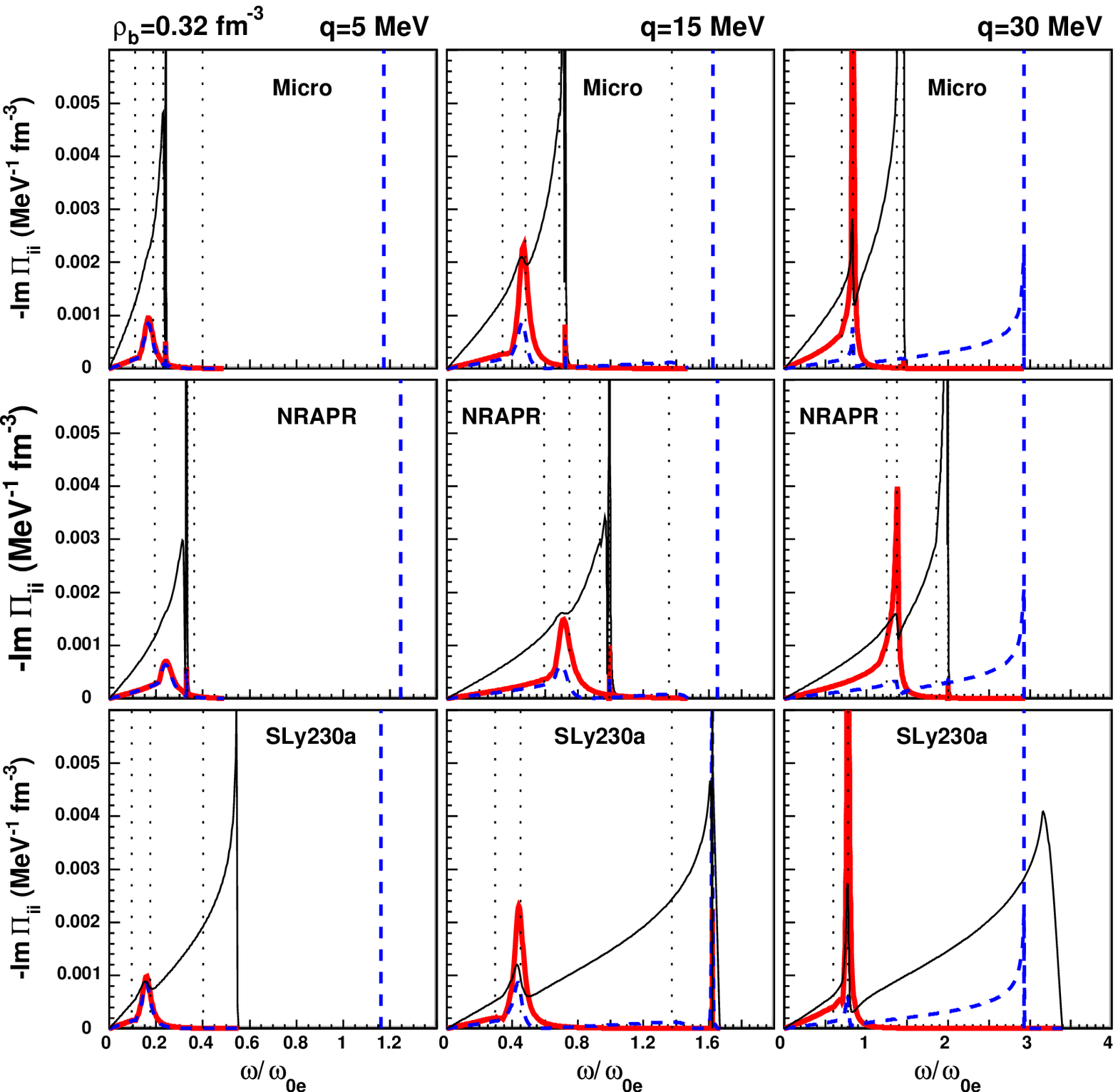}
\caption{
(Color online)
Spectral functions in neutron$+$proton$+$electron matter for selected values of the momentum,
at twice the saturation density and with proton fraction under neutron star matter conditions, for the different nuclear models.
The energy $\omega_{0e}$ is the electron plasmon energy at zero momentum.
The diagonal elements of the spectral function matrix are shown:
$-{\rm Im}\,\Pi^{nn}$ (thin full line),
$-{\rm Im}\,\Pi^{pp}$ (thick full line), and
$-{\rm Im}\,\Pi^{ee}$ (dashed line).
The vertical dotted lines show the energies
at which the determinant $\Delta$ of Eq.~(\ref{eq:Delta}) vanishes.
}
\label{fig15}
\end{center}
\end{figure}

\begin{figure}[h]
\begin{center}
\includegraphics[width=1.0\textwidth]{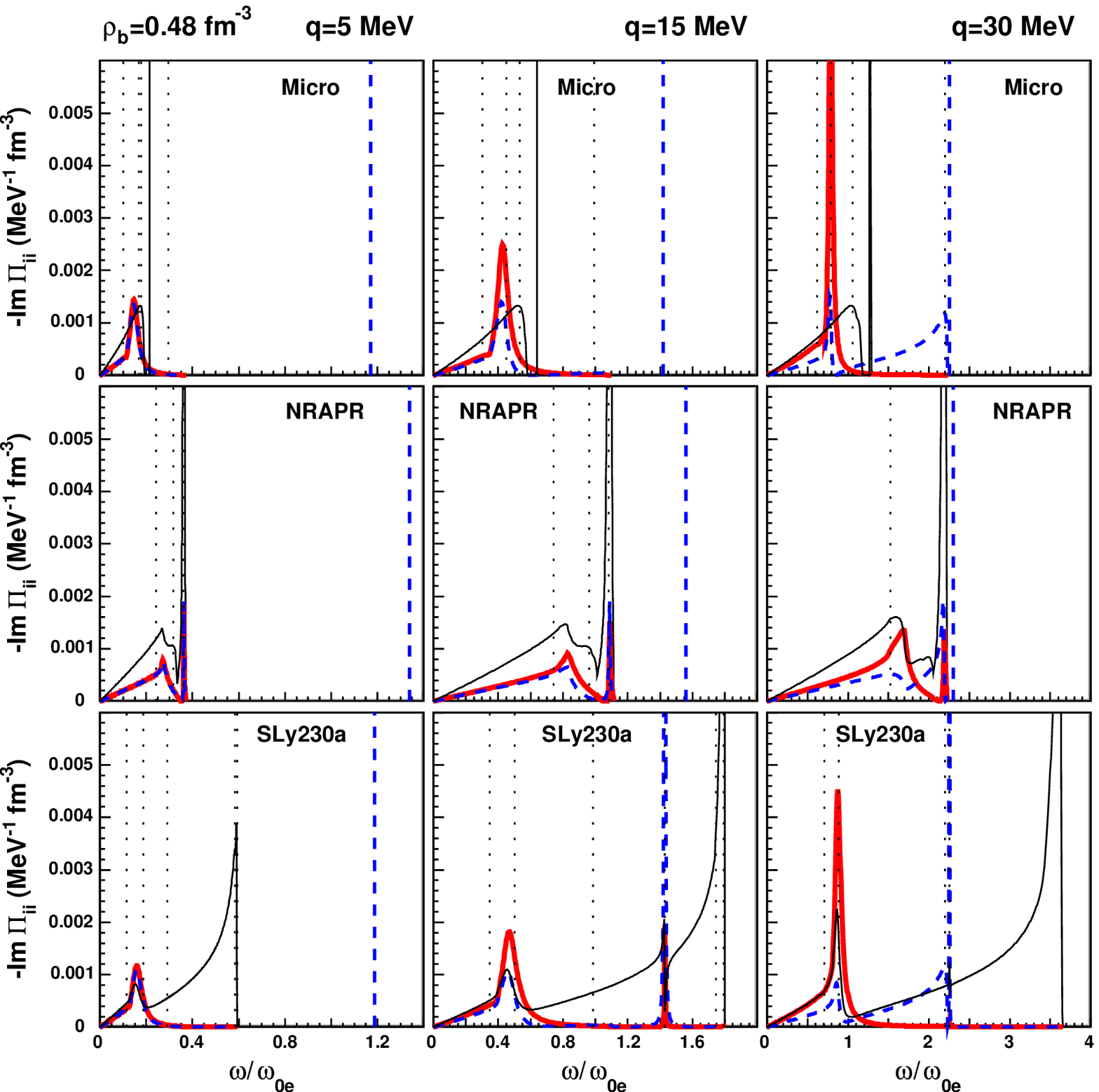}
\caption{
(Color online)
Spectral functions in neutron$+$proton$+$electron matter for selected values of the momentum,
at three times the saturation density and with proton fraction under neutron star matter conditions, for the different nuclear models.
The energy $\omega_{0e}$ is the electron plasmon energy at zero momentum.
The diagonal elements of the spectral function matrix are shown:
$-{\rm Im}\,\Pi^{nn}$ (thin full line),
$-{\rm Im}\,\Pi^{pp}$ (thick full line), and
$-{\rm Im}\,\Pi^{ee}$ (dashed line).
The vertical dotted lines show the energies
at which the determinant $\Delta$ of Eq.~(\ref{eq:Delta}) vanishes.
}
\label{fig16}
\end{center}
\end{figure}

We now study the strength functions $-{\rm Im}\,\Pi^{ii}$ ($i=n,p,e$)
obtained with the different nuclear models, for selected values of the momentum
($q=5$, $15$, $30$ MeV):
they are shown on Figs.~\ref{fig14}, \ref{fig15}, and \ref{fig16},
respectively, for the baryonic density $\rho_0$, $2\rho_0$, and $3\rho_0$.

Let us first comment the behavior of the proton mode.
It is quite similar to the case of the $pe$ model with pure Coulomb coupling:
the mode is always damped, and is mainly a proton mode.
At low momentum, the electron strength function in the peak region is close to the proton one
(electron screening), especially at high density;
however, at higher momentum the proton component of the peak increases,
and becomes much more important than the electron one.
The proton mode is suppressed in two kinds of situations.
When the momentum exceeds the cutoff value of the protonlike branches,
the proton peak becomes a wide bump, which cannot  be identified any more with a collective mode.
This is observed at saturation density, for $q=30$ MeV in the NRAPR and microscopic cases
(see Fig.~\ref{fig14}).
The other case of suppression occurs for NRAPR at $3\rho_0$ (Fig.~\ref{fig16}),
due to the merging of the branches (Fig.~\ref{fig13}).
This effect is due to the low proton effective masses:
this corresponds to a high Fermi velocity $V_{{F}p}$,
which induces a strong coupling with the neutron and electron components.
Finally, we can notice the case of NRAPR at $2\rho_0$ and $q=5$ MeV.
The proton mode appears in the strength function,
and is quite similar to the typical one (Fig.~\ref{fig15});
however, it is not associated with any branch of the dispersion relation,
since at this value of momentum the upper proton branch has not yet appeared
(see Fig.~\ref{fig12}).

Let us now consider the neutron strength function.
In the realistic cases, the imaginary part of ${\rm Im}\,\Pi_0^{n}$ is high when the proton peak occurs,
and although the neutron strength function is affected it does not take the shape of a collective mode.
A real neutron peak associated  with the proton one is only obtained with SLy230a at high density,
when $V_{{F}n}$ is nonphysically high.
However, a specific neutron mode occurs in some cases,
usually indicated by the upper neutronlike branch.
When it appears, this neutron peak is very sharp an high
(exceeding the scale chosen for the present figures).
We should note, however, that the presence of the neutronlike branches
are neither necessary nor sufficient for the presence of such peak.
Let us consider, for instance, the case of density $3\rho_0$
(Figs.~\ref{fig13} and \ref{fig16}).
In the case of NRAPR at $q=30$ MeV, a sharp neutron peak is obtained
although there is no corresponding branch.
On the contrary, in the case of SLy230a at $q=5$ MeV,
the upper neutronlike branch is not associated with a collective mode;
the same kind of behavior has been observed above with the $np$ model.


\section{Conclusion}
\label{sec:conclusion}

The present work is dedicated to the study of normal density modes
in homogeneous $npe$ matter at $\beta$ equilibrium.
Such modes should affect the thermal properties of neutron stars.
We have stressed the behavior of the proton mode:
although it corresponds to a plasmon mode in the jelly model,
the inclusion of electron dynamics reduces it to a soundlike mode,
as was clearly shown in the analysis of the $pe$ system with pure Coulomb coupling.
This effect is also observed for the full $npe$ system,
where the nuclear interaction that couples protons and neutrons is included.

The nuclear quantities of relevance for the study of the density modes in $npe$ matter
(particle-hole interaction and effective mass)
were found to be extremely model dependent:
even the two modern Skyrme forces lead to very different predictions.
This is essentially because of the high isospin-asymmetry values
imposed by the $\beta$ equilibrium in our study:
for such neutron-rich matter, the energy density functional is still poorly constrained.
The effective mass was found to have an important effect
on the dispersion relation obtained in the $(q,\omega)$ plane.
It determines
(i) the Fermi velocities that control the orientation of the different branches,
possibly provoking some merging, and
(ii) the level densities, which are an important quantity for the $q$ extension of the nucleonlike branches.
Note also that for a given repulsive value of $v^{nn}_{\rm{res}}$,
the neutron level density is crucial for the existence or not of the neutronlike branches,
favored by high values of $N_{0n}$.
It should be noted that low values of the nucleon effective masses at high density
are not consistent with a nonrelativistic treatment.
This is a serious limitation of the Skyrme models:
in particular, with SLy230a superluminal values of the neutron Fermi velocity
are reached even before $2\rho_0$.
In any case, we can stress the need to better constrain
the isospin evolution of the nuclear effective masses:
this could be allowed by experimental studies of exotic nuclei
to be performed in the future radioactive-ion-beam facilities,
for instance by the study of the isospin evolution of the level density.
On the other hand, theoretical progress
could improve the determination of the nuclear residual interaction,
possibly through the extraction of Landau parameters with microscopic methods.

An essential conclusion of our study is that the dispersion relation branches in the $(q,\omega)$ plane
are not sufficient to identify collective excitations in neutron-star matter:
they have only an indicative role.
To determine the existence and estimate the damping of such modes,
we have to look for the peaks present in the strength function
related to the imaginary part of the polarization matrix.
Other contributions to the imaginary part,
which we have neglected, come from two-particle$-$two-hole excitations
\cite{Avdee}. However, our aim was to check for each mode if it was
overdamped, and to this purpose the RPA approximation  gives
the correct answer.
Considering the diagonal elements $-{\rm Im}\,\Pi^{ii}$ with $i=n,p,e$,
we have analyzed the density and momentum evolution of the different modes
and identified which components are involved.
In addition to the undamped electronlike plasmon mode,
two nuclear modes have been observed.
At high enough density, the neutron-neutron interaction becomes
repulsive enough to induce a neutron mode,
which appears as a sharp peak in the neutron strength function.
Due to the strong Coulomb repulsion between protons,
a protonlike mode corresponding to the screened proton plasmon
is almost always present.
Its strength increases with $q$ until the region of the proton cutoff momentum,
beyond which the mode is suppressed.
It is interesting to note that a high value of the proton Fermi velocity
can lead to the suppression of this mode by merging into the neutron continuum
(case observed for NRAPR at high density).
The proton collective mode is expected to play a role in the neutrino-emission process
by pair breaking and recombination (PBC) in superfluid matter:
this effect will be discussed in a future work.


\section*{Acknowledgments}
We gratefully thank Dr. H.-J. Schulze for providing us the single-particle potentials as a function of momentum
for different densities and asymmetries
coming from the Brueckner calculations with the v$_{18}$ two-body
interaction and the Urbana IX three-body force, Ref.~\cite{hans}.


\appendix
\section{Residual interaction and energy-density curvature}
\label{ap:parabol}

In the Landau monopolar approximation, the bulk residual interaction
(i.e. the part that is independent of the transferred momentum $q$ of a density fluctuation)
is related to the curvature of the density energy $\mH$
with respect to the single-particle densities $\rho_i$ ($i=n,p$).
In this relation, the role of the effective mass has to be carefully taken into account.
In particular, we show in this appendix that the parabolic approximation for the asymmetry dependence
of the interaction energy breaks down for neutron-rich matter, due to the presence of an effective mass.

The energy density is the sum of a purely kinetic term $\mK$ and an interaction term $\mH_{\rm{int}}$.
In the mean-field approach, it is a functional of the single-particle density,
from which are derived the individual energy levels.
The energy level of a particle of species $i=(n,p)$, bare mass $m_i$ and momentum $k$
takes the general expression:
\beq
\epsilon_i(k,\rho_n,\rho_p)=\frac{k^2}{2m_i}+U_i(k,\rho_n,\rho_p)\;,
\eeq
where $U_i$ is the mean-field potential.
In general, this potential depends on $k$,
which leads to the definition of an effective mass.
The first derivative of the energy density corresponds to the Fermi level:
\be
\frac{\delta\mH}{\delta\rho_i}
&=&\epsilon_{{F}i}(\rho_n,\rho_p)\\
&=&\epsilon_i(k_{{F}i},\rho_n,\rho_p)
=\frac{k_{{F}i}^2}{2m_i}+U_i(k_{{F}i},\rho_n,\rho_p)\;.
\ee
The second derivative reads:
\beq
\frac{\delta^2\mH}{\delta\rho_j\delta\rho_i}
=\frac{\delta\epsilon_{{F}i}}{\delta\rho_j}
=\left(\frac{\delta U_i}{\delta\rho_j}(k_{{F}i},\rho_n,\rho_p)\right)_{k,\rho_{j{\prime}\neq j}={\rm cst}}
+\frac{\delta k_{{F}i}}{\delta\rho_j}
\left(\frac{\delta\epsilon_i}{\delta k}(k_{{F}i},\rho_n,\rho_p)\right)_{\rho_{n,p}={\rm cst}}
\eeq
Introducing the effective mass $m^*_i(\rho_n,\rho_p)$ defined by:
\beq
\label{eq:mstar}
\left(\frac{\delta\epsilon_i}{\delta k}(k_{{F}i},\rho_n,\rho_p)\right)_{\rho_{n,p}={\rm cst}}
= \frac{k_{{F}i}}{m^*_i}
\eeq
and identifying the residual interaction
\beq
v^{ij}_{\rm{res}}(\rho_n,\rho_p)
=\left(\frac{\delta U_i}{\delta\rho_j}(k_{{F}i},\rho_n,\rho_p)\right)_{k,\rho_{j{\prime}\neq j}={\rm cst}}\;,
\eeq
we obtain the relation:
\beq
\frac{\delta^2\mH}{\delta\rho_j\delta\rho_i}
=v^{ij}_{\rm{res}}+\delta_{ij}\frac{\pi^2}{k_{{F}i}m^*_i}\;.
\eeq
Separating the interaction part of the energy density, we have:
\beq
\frac{\delta^2\mH_{\rm{int}}}{\delta\rho_j\delta\rho_i}
=v^{ij}_{\rm{res}}+\delta_{ij}\frac{\pi^2}{k_{{F}i}}\left[\frac{1}{m^*_i}-\frac{1}{m_i}\right]\;.
\eeq
Thus, if the effective mass is different from the bare one,
the residual interaction does not simply reduce to the density curvature of $\mH_{\rm{int}}$.
This is due to the $k$ dependence of the mean-field potential.
Indeed, Eq.~(\ref{eq:mstar}) can be explicited as:
\beq
\frac{1}{m^*_i}
=\frac{1}{m_i}
+\frac{1}{k_{{F}i}}\left(\frac{\delta U_i}{\delta k}(k_{{F}i},\rho_n,\rho_p)\right)_{\rho_{n,p}={\rm cst}}
\eeq
which is the standard definition of the local effective mass at Fermi level.

Let us now consider the proton-proton residual interaction ($i=j=p$) in neutron-rich matter.
At low proton density,
the curvature $\delta^2\mH_{\rm{int}}/\delta\rho_p^2$ diverges as $1/k_{{F}p}$
because of the effective mass.
Treating the isospin dependence of $\mH_{\rm{int}}$ in the parabolic approximation
amounts to ignoring this effect, and leads to a large underestimation of $v^{pp}_{\rm{res}}$
in the case $m^*_p<m_p$.

\end{document}